\documentclass[12pt]{article}

\usepackage{amsmath,amsfonts,mathtools}
\usepackage[a4paper,margin=1.0in]{geometry} 
\usepackage{graphicx,graphics,subfigure}
\usepackage{microtype}
\usepackage[colorlinks,linktocpage,citecolor=DarkGreen,linkcolor=DarkRed,urlcolor=DarkBlue,pdfborder={0 0 0},hyperfootnotes=false]{hyperref}
\usepackage{txfonts}
\usepackage{setspace}
\usepackage[square,comma,numbers,sort&compress]{natbib}
\usepackage[utf8]{inputenc}
\usepackage[english]{babel}
\usepackage{float}
\usepackage[margin=10pt,font={small,stretch=1.4},labelfont=bf]{caption}
\usepackage[usenames,dvipsnames,svgnames,table]{xcolor}
\usepackage{epstopdf}
\usepackage{enumerate}
\begin{document}

\begin{titlepage}
\hypersetup{pageanchor=false}
\begin{center}

\vspace*{0.1mm}

\begin{spacing}{2.5}
{\LARGE \textbf{Tensor Perturbations and Thick Branes \\ in Higher-dimensional \texorpdfstring{$f(R)$}{f(R)} Gravity}}
\end{spacing}

\vskip 10mm

\begin{spacing}{1.5}
\large
Zheng-Quan Cui, 
Zi-Chao Lin, 
Jun-Jie Wan, \\ 
Yu-Xiao Liu, and  
Li Zhao\footnote{Email: \texttt{lizhao@lzu.edu.cn}, corresponding author}
\end{spacing}

\vskip 5mm

{
\emph{Joint Research Center for Physics, Lanzhou University and Qinghai Normal University, \\ Lanzhou 730000 and Xining 810000, China\\}
\emph{Institute of Theoretical Physics $\&$ Research Center of Gravitation, Lanzhou University,\\
                            Lanzhou 730000, China}\\
\emph{Key Laboratory for Magnetism and Magnetic Materials of the MOE, Lanzhou University,\\
                            Lanzhou 730000, China}\\
\emph{Lanzhou Center for Theoretical Physics, Lanzhou University, \\
                          Lanzhou 730000, China}
}

\vskip 5mm


\end{center}

\vskip 10mm

\begin{center}
{\textbf{\textsc{Abstract}}}
\end{center}
We study brane worlds in an anisotropic higher-dimensional spacetime within the context of $f(R)$ gravity. Firstly, we demonstrate that this spacetime with a concrete metric ansatz is stable against linear tensor perturbations under certain conditions. Moreover, the Kaluza-Klein modes of the graviton are analyzed. Secondly, we investigate thick brane solutions in six dimensions and their properties. We further exhibit two sets of solutions for thick branes. At last, the effective potential of the Kaluza-Klein modes of the graviton is discussed for the two solved $f(R)$ models in higher dimensions.

\vskip 5mm


%

\end{titlepage}

\newpage
\hypersetup{pageanchor=true}
\pagenumbering{arabic} 

\begin{spacing}{1.2}
\tableofcontents
\end{spacing}

\section{Introduction}

General relativity (GR) is successful in many fields but leaves a few issues up in the air. Both phenomenological and theoretical investigations reveal modifications to GR under certain circumstances. Higher curvature terms may serve as these modifications and many investigations start from them but out of various considerations. However, they may lead to higher derivatives in equations of motion rather than second ones, which are generally difficult to be solved. $f(R)$ theory of gravitation~\cite{Sotiriou:2010sf,De Felice:2010dt}, one of the simplest higher derivative generalizations of GR, is possible to be analytically explored in some situations. Although $f(R)$ theory is an effective theory, its success in many fields has attracted extensive attention.

On the other hand, the brane world scenario provides an alternative approach to address outstanding issues in four dimensions. Remarkably, the brane world model with a warped extra dimension pioneered by Randall and Sundrum~\cite{Randall:1999rsa} has drawn wide attention since it exhibits the possibility of an infinite fifth dimension without violating known experiments of gravitation. The general brane world sum rules indicate some particular classes of five-dimensional brane world models in $f(R)$ gravity~\cite{da Silva:2011sd}. Moreover, due to the inevitable appearance of higher derivatives in equations of motion, investigation on junction conditions of $f(R)$ gravity in the brane world scenarios opens the possibility of a new class of thin brane  solutions~\cite{Deruelle:2007dss,Borzou:2009bssy,Balcerzak:2008bd,Balcerzak:2009bd,Balcerzak:2011bd,Carames:2013cgs,Parry:2005ppd}.

Thick branes, domain walls~\cite{Rubakov:1983rsd} with a warp spacetime background, can preferably circumvent the requirements of the junction conditions and naturally remove the divergence of curvature.
Recent reviews including $f(R)$ thick branes refer to Refs.~\cite{Maartens:2010gt,Dzhunushaliev:2010dfm,Liu:2017l}. To avoid solving higher derivative equations, $f(R)$ gravity within the context of brane worlds has been considered via the conformal equivalence between $f(R)$ theory and GR with a scalar field~\cite{Parry:2005ppd,Zhong:2016zl}. However, this approach needs to transform the solutions of second derivative equations of motion back to the physical frame. Thick branes in the higher-order frame were explored by numerical or approximate approaches~\cite{Dzhunushaliev:2010dfkk,Dzhunushaliev:2019dfno,Dzhunushaliev:2019dfs,Afonso:2007abmp}. In Ref.~\cite{Afonso:2007abmp}, thick branes with constant curvature were investigated, but these solutions may not lead to the usual four-dimensional gravitation. Nontrivial analytical thick brane solutions with nonconstant curvature in $f(R)$ theory were first investigated in Ref.~\cite{Liu:2011lzzl} and further considered in Refs.~\cite{Liu:2012llw,Bazeia:2013bmps,Bazeia:2014blmps,Xu:2015xzyl,Bazeia:2015blm,Bazeia:2015bllmo,Zhong:2016zl,Bazeia:2014blmor,Gu:2015ggyl,Gu:2017gzyl,Gu:2018glz,Cui:2018clgz}.  Unlike the approach by introducing background scalar fields to construct thick branes, pure geometric frameworks (without background matter fields) were taken into account in Refs.~\cite{Liu:2012llw,Dzhunushaliev:2010dfkk,Zhong:2016zl,Wang:2019wgfx,Dzhunushaliev:2019dfno,Dzhunushaliev:2019dfs}.

Most of the related works are investigated in five-dimensional spacetime. It is worth noting that the known thick $f(R)$ branes in higher spacetime dimensions~\cite{Liu:2012llw,Dzhunushaliev:2019dfno,Dzhunushaliev:2019dfs} are constructed in pure gravity without background scalar fields. However, to localize a bulk fermion field on the brane, the Yukawa coupling between the fermion field and a background scalar field is usually needed. Moreover, in five dimensions, the pure gravitational trapping mechanism of vector fields remains problematic and there is no remarkable proposal for the fermion mass hierarchy in the Standard Model. These issues may employ higher dimensions to address. Our goal is to investigate brane solutions with nonconstant curvature in the context of six-dimensional $f(R)$ gravity with a real scalar field.


Before exploring solutions in a background spacetime, we should first consider its perturbation stability. Considering four-dimensional Poincar\'{e} symmetry, the decomposition of perturbations will give rise to massless and massive graviton Kaluza-Klein (KK) modes. A localized massless graviton KK mode contributes to the four-dimensional Newtonian potential and the massive ones lead to corrections to the Newtonian potential. Gravitational resonant modes, a class of massive KK modes, have been studied for various solved $f(R)$ models in brane world scenarios~\cite{Xu:2015xzyl,Yu:2016yzgl}. More complete analyses on perturbations with extra spatial dimensions in the context of GR have been achieved in Refs.~\cite{Csaki:2000cehs,Kobayashi:2002kks,Giovannini:2001gg,Giovannini:2002ga,Giovannini:2002gb,Giovannini:2003g,Giovannini:1997g}. The linear stability of the tensor perturbation of $f(R)$ brane models was firstly investigated in Ref.~\cite{Zhong:2011zly}. Other related investigations can also be seen in Refs.~\cite{Cui:2018clgz,Zhong:2016zl,Zhong:2017zl,Bazeia:2015bllmo,Gu:2017gzyl,Gu:2015ggyl,Gu:2018glz}. Scalar perturbations within the context of pure $f(R)$ gravity have been elaborated by the transformation of the $f(R)$ theory to a scalar–tensor theory~\cite{Zhong:2016zl,Gu:2018glz} and directly studied in the higher-order frame~\cite{Zhong:2017zl}. However, scalar perturbations in $f(R)$ gravity with background scalar fields are difficult to be analyzed because of the coupling between the scalar modes of the metric perturbations and background scalar fields, and the resolution was given in Ref.~\cite{Chen:2017cgl}. For our research, we investigate the linear stability of higher-dimensional $f(R)$ gravity in brane world scenarios against tensor perturbations.


This paper is organized as follows. In section~\ref{sec:model}, we consider $f(R)$ gravity in a $D$-dimensional spacetime and give the equations of motion under a concrete metric describing flat branes. In section~\ref{sec:stability}, we investigate the linear stability of this background spacetime against tensor perturbations. In section~\ref{sec:KKmodes}, the graviton KK modes are discussed. In section~\ref{sec:6dmodel}, we seek for analytic solutions of thick branes in a six-dimensional bulk spacetime. Exact solutions of the scalar field with a domain wall configuration are studied. In section~\ref{sec:epotential}, we discuss the effective potential in solved $f(R)$ models in higher dimensions. Finally, discussions and conclusions are given in section~\ref{sec:conclusions}.

Throughout the paper, capital Latin letters $M,N,\ldots$ represent the $D\,(=4+1+d)$-dimensional coordinate indices running over $0, 1, 2, 3, 5, 6, \ldots$, lower-case Greek letters $\mu,\nu,\ldots$ represent the four-dimensional Minkowski coordinate indices running over $0, 1, 2, 3$, and lower-case Latin letters $i,j,\ldots$ represent the extra-spatial Euclidean coordinate indices running over $6, 7, \ldots$. We use the definitions $R^P_{MQN}=\partial_Q\Gamma^P_{MN}-\partial_N\Gamma^P_{MQ}+\Gamma^P_{QL}\Gamma^L_{MN}-\Gamma^P_{NL}\Gamma^L_{MQ}$ and $R_{MN}=R^L_{MLN}$. The metric signature is $(-,+,+,\cdots,+)$.


\section{\texorpdfstring{$f(R)$}{f(R)} Gravity in \texorpdfstring{$D$}{D}-dimensional Spacetime}
\label{sec:model}

We start with the following $D$-dimensional action within the context of $f(R)$ gravity (for reviews see Refs.~\cite{Sotiriou:2010sf,De Felice:2010dt}),
\begin{equation}
S=\frac{1}{2\kappa_D^2}\int\mathrm{d}^Dx\sqrt{-g^{(D)}}f(R)+S_\mathrm{m}\,,\label{actionall}
\end{equation}
where $g^{(D)}\equiv\mathrm{det}\,g_{MN}$, and $\kappa_D^2=8\pi G_{\mathrm{N}}^{(D)}=1/M_{(D)}^{D-2}$ with $G_{\mathrm{N}}^{(D)}$ the $D$-dimensional gravitational constant and $M_{(D)}$ the $D$-dimensional fundamental scale. The variation of the action~\eqref{actionall} with respect to the metric $g_{MN}$ yields the following field equation
\begin{equation}\label{eq:fieldeq1}
f_{R} R_{M N}-\frac{1}{2} f g_{M N}-\nabla_{M} \nabla_{N} f_{R}+g_{M N} \square^{(D)} f_{R}=\kappa_D^{2} T_{M N}\,,
\end{equation}
where $f_{R}$ denotes $\mathrm{d} f/\mathrm{d} R$, $\square^{(D)}=g^{AB}\nabla_{A}\nabla_{B}$ is the $D$-dimensional d'Alembert operator, and $T_{MN}=-\frac{2}{\sqrt{-g^{(D)}}}\frac{\delta S_\mathrm{m}}{\delta g^{MN}}$ is the energy-momentum tensor.

Specifically, a particular example of an anisotropic $(4+1+d)$-dimensional spacetime $\mathcal{M}_4\times\mathcal{R}_1\times\mathcal{E}_d$ is considered, where $\mathcal{M}_4$ is a four-dimensional Minkowski manifold, $\mathcal{R}_1$ is a special manifold with a noncompact extra spatial dimension, and $\mathcal{E}_d$ is a Euclidean manifold with $d$ extra spatial dimensions. In this paper, we are interested in a four-dimensional static flat spacetime embedded in the $(4+1+d)$-dimensional bulk, which takes the form
\begin{equation}\label{linee}
\mathrm{d}s^2=\mathrm{e}^{2A(y)}\eta_{\mu\nu}\mathrm{d}x^\mu\mathrm{d}x^\nu+\mathrm{d}y^2+
\mathrm{e}^{2B(y)}\delta_{ij}\mathrm{d}\hat{x}^i\mathrm{d}\hat{x}^j\,.
\end{equation}
Here $\mathrm{e}^{A(y)}$ and $\mathrm{e}^{B(y)}$ are warp factors which give rise to the warped geometry, $\eta_{\mu\nu}$ and $\delta_{ij}$ are metrics in the $\mathcal{M}_4$ and the $\mathcal{E}_d$, respectively, and $y=x^5$ is the spacial extra-dimensional coordinate. With the coordinate transformations $\mathrm{d}z=\mathrm{e}^{-A(y)}\mathrm{d}y$ and $\mathrm{d}w^i=\mathrm{e}^{B(y)-A(y)}\mathrm{d}\hat{x}^i$, the above metric can be rewritten as
\begin{equation}\label{lineec}
\mathrm{d}s^2=\mathrm{e}^{2A(z)}\left(\eta_{\mu\nu}\mathrm{d}x^\mu\mathrm{d}x^\nu+\mathrm{d}z^2+
\delta_{ij}\mathrm{d}w^i\mathrm{d}w^j\right)\,.
\end{equation}
Denoting $a(y)=\mathrm{e}^{A(y)}$ and $b(y)=\mathrm{e}^{B(y)}$ and using the metric ansatz~\eqref{linee}, Eq.~\eqref{eq:fieldeq1} is reduced to
\begin{subequations}\label{eq:eq1MN}
\begin{align}
&(\mu,\nu): &  \frac{1}{2}f+\left(3\frac{a'^{2}}{a^{2}}+\frac{a''}{a}+d\frac{a'b'}{ab}\right)f_{R}&
-\left(3\frac{a'}{a}+d\frac{b'}{b}\right)f_{R}'-f_{R}''=
-\frac{\kappa_D^{2}}{4}\frac{\eta^{\mu\nu}}{a^2}T_{\mu\nu}\,,\label{eq:eq1compuv}\\
&(y,y): &  \frac{1}{2}f+\left(4 \frac{a''}{a}+d\frac{b''}{b}\right)f_{R}&-\left(4 \frac{a'}{a}+d\frac{b'}{b}\right)f_{R}'=-\kappa_D^{2}T^5_{\;5}\,,\label{eq:eq1comp55}\\
&(i,j): &  \frac{1}{2}f+\left[(d-1)\frac{b'^2}{b^2}+\frac{b''}{b}+4 \frac{a'b'}{ab}\right]f_{R}&
-\left[4\frac{a'}{a}+(d-1)\frac{b'}{b}\right]f_{R}^{\prime}-f_{R}^{\prime\prime}
=-\frac{\kappa_D^{2}}{d}\frac{\delta^{ij}}{b^2}T_{ij}\,,\label{eq:eq1compij}
\end{align}
\end{subequations}
where the prime denotes the derivative with respect to the extra-dimensional coordinate $y$. To explore the stability of this configuration and effective gravitation in the $\mathcal{M}_4$, we would like to examine perturbations concerning this background spacetime.

\section{Linear Stability}
\label{sec:stability}

In $D$-dimensional spacetime, pure $f(R)=R+\alpha R^n$ gravity (without matter) has been considered for thick brane solutions~\cite{Dzhunushaliev:2019dfno,Dzhunushaliev:2019dfs}. Nevertheless, the perturbative stability of spacetime is still unknown. We mainly focus on the linear stability of spacetime under perturbations and
begin with gravitational perturbations for general $f(R)$ gravity in $D$-dimensional spacetime (for general perturbations see appendix~\ref{appendix1}). For a deformation of Eq.~\eqref{eq:fieldeq1}, the linearized field equation is
\begin{align}\label{eq:perturbedfieldeq1}
\delta f_{R} \,G_{MN}+f_{R}\delta G_{MN}=
\frac{1}{2}&\left[\left(\delta f-\delta R\,f_{R}-R\delta f_{R}\right)g_{MN}
+\left(f-Rf_{R}\right)\delta g_{MN}\right] \nonumber  \\ & +\delta\left(\nabla_M\nabla_Nf_{R}\right)-\delta\left(g_{MN}\square^{(D)}f_{R}\right)
+\kappa_D^{2} \delta T_{M N} \,,
\end{align}
where the $\delta$ denotes a linear perturbation. This equation can also be derived by variation with respect to the action including quadratic terms of perturbations.

With the help of the expansions of $\nabla_M\nabla_Nf_{R}$ and $g_{MN}\nabla_A\nabla^Af_{R}$:
\begin{subequations}\label{relations}
\begin{align}
\nabla_M\nabla_Nf_{R}&=\left(\partial_M\partial_N-\Gamma^P_{NM}\partial_P\right)f_{R}\,,\\
g_{MN}\square^{(D)}f_{R}&=g_{MN}\nabla_A\nabla^Af_{R}=g_{MN}g^{AB}\left(\nabla_A\nabla_Bf_{R}\right)\,,
\end{align}
\end{subequations}
one can write the two terms $\delta\left(\nabla_M\nabla_Nf_{R}\right)$ and $\delta\left(g_{MN}\nabla_A\nabla^Af_{R}\right)$ on the right hand side of Eq.~\eqref{eq:perturbedfieldeq1} as
\begin{subequations}
\begin{align}
\delta\left(\nabla_M\nabla_Nf_{R}\right)&=
\left(\partial_M\partial_N-\Gamma^P_{NM}\partial_P\right)\delta f_{R}-\delta\Gamma^P_{NM}\partial_Pf_{R}\,,\label{expanddelta1}\\
\delta\left(g_{MN}\square^{(D)}f_{R}\right)&=\delta\left(g_{MN}g^{AB}\nabla_A\nabla_Bf_{R}\right) \nonumber \\
&=\delta g_{MN}\square^{(D)}f_{R}+g_{MN}\delta g^{AB}\left(\nabla_A\nabla_Bf_{R}\right)+ g_{MN}g^{AB}\delta\left(\nabla_A\nabla_Bf_{R}\right)\,.\label{expanddelta2}
\end{align}
\end{subequations}

We investigate perturbations under the background metric~\eqref{linee} in the following. For an observer localized on the $\mathcal{M}_4$, the perturbations of the metric~\eqref{linee} and bulk fields can be decomposed into the transverse-traceless (TT) tensor mode, transverse vector modes, and scalar modes according to the four-dimensional Lorentz transformation. The perturbations of the metric couple to the perturbations of bulk fields. Each type of these perturbation modes obeys independent equations of motion at the linearized level~\cite{Giovannini:1997g,Giovannini:2001gg,Giovannini:2002ga,Giovannini:2002gb,Giovannini:2003g}. For instance, the perturbations of a scalar field in the bulk does not appear in the equation of motion of the tensor mode. In Refs.~\cite{Giovannini:1997g,Giovannini:2001gg,Giovannini:2002ga,Giovannini:2002gb,Giovannini:2003g}, the linear perturbation stability analysis was applied for the line element~\eqref{lineec} in GR. Taking a step further, we examine the following linear tensor perturbations in the context of $f(R)$ gravity:
\begin{equation}\label{perturbation}
\mathrm{d}s^2=a^2(y)(\eta_{\mu\nu}+h_{\mu\nu})\mathrm{d}x^\mu\mathrm{d}x^\nu+\mathrm{d}y^2+
b^2(y)\delta_{ij}\mathrm{d}\hat{x}^i\mathrm{d}\hat{x}^j\,,
\end{equation}
where $h_{\mu\nu}$ represents the TT tensor mode obeying
\begin{align}\label{TTcondition}
\partial_\mu h^\mu_{\nu}=0\,,\quad h=\eta^{\mu\nu}h_{\mu\nu}=0\,.
\end{align}
Some fundamental quantities of spacetime under such a perturbation can be calculated accordingly. They are collected in appendix~\ref{appendix2}.


The $(\mu,\nu)$ components of $\delta\left(\nabla_M\nabla_Nf_{R}\right)$ and $\delta\left(g_{MN}\square^{(D)}f_{R}\right)$ under the perturbations~\eqref{perturbation} are calculated as
\begin{subequations}\label{TTf}
\begin{align}
\delta\left(\nabla_\mu \nabla_\nu f_{R}\right)=&a^2\left(\frac{a^{\prime}}{a}+\frac{1}{2}f_{R}^{\prime} \partial_y\right)h_{\mu \nu}\,, \\
\delta\left(g_{\mu\nu}\square^{(D)}f_{R}\right)=&a^{2}\left[\left(4 \frac{a^{\prime}}{a}+d \frac{b^{\prime}}{b}\right) f_{R}^{\prime}+f_{R}^{\prime \prime}\right]h_{\mu \nu}\,.
\end{align}
\end{subequations}
Plugging the expression of $\delta G_{\mu\nu}$~\eqref{deltaGmunu} (see appendix~\ref{appendix2}) into Eq.~\eqref{eq:perturbedfieldeq1}, and considering $\delta\left(\nabla_\mu \nabla_\nu f_{R}\right)$ and $\delta\left(g_{\mu\nu}\square^{(D)}f_{R}\right)$ in Eq.~\eqref{TTf}, the $(\mu,\nu)$ component of Eq.~\eqref{eq:perturbedfieldeq1} is worked out to be
\begin{align}
&\left[\frac{1}{2}f+\left(3\frac{a'^{2}}{a^{2}}+\frac{a''}{a}+d\frac{a'b'}{ab}\right)f_{R}
-\left(3\frac{a'}{a}+d\frac{b'}{b}\right)f_{R}'-f_{R}''
+\frac{\kappa_D^{2}}{4}\frac{\eta^{\alpha\beta}}{a^2}T_{\alpha\beta}\right] h_{\mu \nu} \nonumber \\
&+\frac{1}{2}f_{R}\left[\frac{1}{a^2}\square^{(4)} h_{\mu \nu}+\frac{1}{b^2}\hat{\Delta}^{(d)}h_{\mu \nu}+
\left(4 \frac{a'}{a}+d\frac{b'}{b}\right) h'_{\mu \nu}+ h''_{\mu\nu}\right]
+\frac{1}{2} f_{R}^{\prime} h'_{\mu \nu}=0\,.  \label{munucomponent}
\end{align}
Here, $\square^{(4)}=\eta^{\mu\nu}\partial_\mu\partial_\nu$ and $\hat{\Delta}^{(d)}=\delta^{ij}\partial_i\partial_j$ are the d'Alembert operator in the $\mathcal{M}_4$ and the Laplace operator in the $\mathcal{E}_d$, respectively. Taking the background equation~\eqref{eq:eq1compuv} into account, the main perturbed equation~\eqref{munucomponent} for the TT tensor mode is reduced to
\begin{align}\label{eq:perturbedequvend}
\left[\frac{1}{a^2}\square^{(4)}
      +\frac{1}{b^2}\hat{\Delta}^{(d)}
      +\left(4 \frac{\partial_y a}{a}+d\frac{\partial_y b}{b}\right) \partial_y
      + \partial_{y} \partial_{y}
\right] h_{\mu \nu}
+\frac{\partial_y f_{R}}{f_{R}}\, \partial_{y} h_{\mu \nu}=0\,.
\end{align}
Since
\begin{equation}
\square^{(D)} = \frac{1}{a^2}\square^{(4)}
      +\frac{1}{b^2}\hat{\Delta}^{(d)}
      +\left(4 \frac{\partial_y a}{a}+d\frac{\partial_y b}{b}\right) \partial_y
      + \partial_{y} \partial_{y},
\end{equation}
for the curved background~\eqref{linee}, the above equation~\eqref{eq:perturbedequvend} can be written as
\begin{equation}
\square^{(D)} h_{\mu \nu}=-\left(\partial_y \ln f_{R}\right)\, \partial_{y} h_{\mu \nu}\,.
\end{equation}
In the coordinates $\left(x^{\mu}, z, w^i\right)$, Eq.~\eqref{eq:perturbedequvend} turns into
\begin{equation}
\left[\partial_z^{2}+\partial_z\ln \left(a^3b^df_R\right)\,\partial_z+\Delta^{(d)}+\square^{(4)}\right]h_{\mu\nu}=0\,,
\end{equation}
where $\Delta^{(d)}=\delta^{ij}\partial_{w^i}\partial_{w^j}$ represents the Laplace operator in the $\mathcal{E}_d$ regarding the transformed extra-dimensional coordinates $w^i$.

Next, we perform the following separation of variables
\begin{align}\label{KKdecom}
  h_{\mu\nu}\left(x^{\mu},z,w^i\right)=\epsilon_{\mu\nu}\left(x^{\mu}\right)
  \left[a^3(z)b^d(z)f_{R}(z)\right]^{-1/2}\psi\left(z\right)\xi(w^i)\,.
\end{align}
Then we obtain the Klein-Gordon equation for the four-dimensional part $\epsilon_{\mu\nu}(x^{\mu})$:
\begin{equation}\label{eq:KG1}
\left(\square^{(4)}-m^2\right)\epsilon_{\mu\nu}\left(x^{\mu}\right)=0\,,
\end{equation}
the Schr\"{o}dinger-like equation for the fifth-dimensional spatial part $\psi(z)$:
\begin{equation}\label{eq:Schrodinger}
\left[-\partial_z^2+W(z)\right]\psi(z)=\left(m^2-l^2\right)\psi(z)\,,
\end{equation}
and the Helmholtz equation for the extra $d$-dimensional spatial part $\xi(w^i)$:
\begin{equation}\label{eq:KG2}
\left(\Delta^{(d)}+l^2\right)\xi(w^i)=0\,.
\end{equation}
Here the effective potential $W(z)$ is of the following form
\begin{align}
W(z)=&\frac{1}{4}\left(3\frac{\left(\partial_{z}a\right)^2}{a^2}
+d(d-2)\frac{\left(\partial_{z}b\right)^2}{b^2}
-\frac{\left(\partial_{z}f_{R}\right)^2}{f_{R}^2}\right)+
\frac{1}{2}\left(3d\frac{\partial_{z}a}{a}\frac{\partial_{z}b}{b}+
3\frac{\partial_{z}a}{a}\frac{\partial_{z}f_{R}}{f_{R}}
+d\frac{\partial_{z}b}{b}\frac{\partial_{z}f_{R}}{f_{R}}\right) \nonumber \\
&+\frac{1}{2}\left(3\frac{\partial_{z}\partial_{z}a}{a}+d\frac{\partial_{z}\partial_{z}b}{b}
+\frac{\partial_{z}\partial_{z}f_{R}}{f_{R}}\right) \nonumber \\
=&\Omega^2+\partial_z\Omega \label{epotential}
\end{align}
with
\begin{equation}\label{Omega}
\Omega=\frac{1}{2}\partial_z \ln \left(a^3b^df_R\right)\,.
\end{equation}
Both $m^2<0$ and $l^2<0$ are not physically reasonable, they will lead to the solution~\eqref{KKdecom} either evolving exponentially in time or increasing exponentially in space. Essentially, the Schr\"{o}dinger-like equation~\eqref{eq:Schrodinger} can be factorized as a supersymmetric quantum mechanics form
\begin{equation}\label{eq:ssform}
 \mathcal{Q}\,\mathcal{Q}^{\dagger}\psi(z)=\left(m^2-l^2\right)\psi(z)
\end{equation}
with
\begin{equation}
\mathcal{Q}=\partial_z+\Omega\,,\quad \mathcal{Q}^{\dagger}=-\partial_z+\Omega\,,
\end{equation}
which ensures $m^2-l^2\geqslant 0$. We conclude that this system is stable under tensor perturbations. However, the condition $m^2-l^2\geqslant 0$ does not ensure that there are no lower energy states which are apparent tachyons. If one demands that apparent tachyon states are absent in the $\mathcal{M}_4$, $m^2\geqslant 0$ should be satisfied. The two conditions $m^2-l^2\geqslant 0$ and $m^2\geqslant 0$ are significant for the stability. We give a brief discussion for several cases:
\begin{itemize}
  \item there are no any apparent tachyon states in the $\mathcal{M}_4$ if $m^2-l^2\geqslant 0$ and $m^2\geqslant 0$ are satisfied;
  \item the case $m^2-l^2\geqslant 0$ and $m^2<0$ will result in some apparent tachyon states of the graviton in the $\mathcal{M}_4$;
  \item it is worth stressing that $m^2=0$ and $l^2=0$ will lead to a zero energy state which stands for the graviton zero mode; however, a zero energy state does not indicate $m^2=0$ and $l^2=0$ for the case of $m^2=l^2$.
\end{itemize}

Until now, the above analyses of tensor perturbations are independent of the explicit background spacetime. It is worth noting that the above results are applicable for $d\geqslant 0$. For $d=0$, namely the five-dimensional case, it was reported in Ref.~\cite{Zhong:2011zly}. In the case of $d>0$, the effective potential compared with the one with $d=0$ will be corrected due to the presence of the warp factor $b$. Moreover, the above derivation can also be done with the line element~\eqref{lineec}. Some related works in the context of GR can be seen in Refs.~\cite{Giovannini:1997g,Giovannini:2001gg,Giovannini:2002ga,Giovannini:2002gb,Giovannini:2003g}.

Furthermore, $f(R)$ gravity suffers from instabilities on account of ghosts or tachyons~\cite{Sotiriou:2010sf,De Felice:2010dt}. Stability conditions, i.e., the ghost-free condition $\frac{\mathrm{d}f(R)}{\mathrm{d}R}>0$ and the tachyon-free condition $\frac{\mathrm{d}^2f(R)}{\mathrm{d}R^2}>0$, should be fulfill physically to make sure that $f(R)$ theory is viable.  Nevertheless, the ghost-free condition is a local condition only for short wavelength modes. There is the possibility of the existence of other linear instabilities in the spacetime considered here.

\section{The KK Modes of the Graviton}
\label{sec:KKmodes}

In what follows, we discuss the KK modes of the tensor perturbations which will help us to analyze the four-dimensional effective theory. For a four-dimensional observer in the $\mathcal{M}_4$, the KK modes of the tensor perturbations reflect the configuration of extra dimensions. We use $h_{\mu\nu}^{(ml)}(x^M)$ to denote the KK modes with $\xi_{(l)}(w^i)$ and $\psi_{(ml)}\left(z\right)$ their $w^i$- and $z$-coordinate parts. If a KK mode is normalizable, it corresponds to a four-dimensional graviton. There are a series of modes with $m^2-l^2\geqslant 0$ in terms of the Schr\"{o}dinger-like equation~\eqref{eq:Schrodinger}. Regardless of the value of $l^2$, these modes are called massless if $m^2=0$ and massive if $m^2\neq0$. One can find a special set of modes $\psi_{m^2=l^2}(z)$ with $m^2=l^2$. If we impose periodic boundary conditions on the $\mathcal{E}_d$, the condition $l^2\geqslant 0$ will be satisfied. There is a special mode $\psi_{(00)}(z)$ with $m^2=l^2=0$ corresponding to the graviton zero mode.

We clarify the condition that one can obtain a four-dimensional effective gravitational theory from the perspective of action reduction. To quadratic order in tensor perturbations, the gravitational part of the action~\eqref{actionall} is
\begin{align}
  S_{\mathrm{g}}=\frac{1}{2\kappa_D^2}\int\mathrm{d}^Dx\left[f(R)\;\delta^{(2)}\sqrt{-g^{(D)}}
  +\sqrt{-g^{(D)}}\left(f_R\delta^{(2)}R+\frac{1}{2}f_{RR}(\delta R)^2\right)+ f_R\delta R\;\delta\sqrt{-g^{(D)}}\right]\,,
\end{align}
where $\delta^{(2)}$ denotes second-order quantities in tensor perturbations and $f_{RR}$ denotes $\mathrm{d} f^2/\mathrm{d} R^2$. Taking the TT condition~\eqref{TTcondition} and the separation of variables~\eqref{KKdecom} in the coordinates $\left(x^{\mu}, z, w^i\right)$ into account, one may obtain a four-dimensional effective theory with the Einstein-Hilbert action and some additional terms:
\begin{align}\label{eaction}
  S_{\textrm{eff}}\supset\frac{M_{(D)}^{D-2}}{2}\int\mathrm{d}^4x \;\partial^\alpha \epsilon^{\rho\lambda}\left(x^{\mu}\right)\partial_\alpha \epsilon_{\rho\lambda}\left(x^{\mu}\right) \int\mathrm{d}^d w\;\xi^2(w^i) \int\mathrm{d}z\;\psi^2(z)\,.
\end{align}
Such a four-dimensional effective action is anticipated to describe gravitation observed experimentally. The four-dimensional effective Planck mass $M_{\textrm{Pl}}$ is expressed as follows
\begin{align}\label{Pl}
  M_{\textrm{Pl}}^2=M_{(D)}^{D-2}\int_{\mathcal{E}_d}\mathrm{d}^d w\;\xi^2(w^i)
  \int_{-\infty}^{+\infty}\mathrm{d}z\;\psi^2(z)\,,
\end{align}
where $\int_{\mathcal{E}_d}\mathrm{d}^d w\equiv V$ is the volume of the $\mathcal{E}_d$. Since $\xi(w^i)$ is a plane wave solution, one only needs that the integral domain is finite. If the volume $V$ is finite, an arbitrary KK mode is normalizable if and only if the following normalization condition is satisfied
\begin{align}\label{normalizationc}
  \int_{-\infty}^{+\infty}\psi^2(z) \;\mathrm{d}z=\int_{-\infty}^{+\infty}\psi^2(z(y))\; a^{-1}(y)\;\mathrm{d}y<\infty\,.
\end{align}
As a consequence, one can obtain a four-dimensional effective action from the higher-dimensional $f(R)$ gravity~\eqref{actionall} under the metric ansatz~\eqref{linee}. Specifically, the localized massless zero mode results in a four-dimensional GR and hence the four-dimensional Newtonian potential. Therefore, the zero mode must satisfy the normalization condition~\eqref{normalizationc}. Regarding to the massive KK modes, they should not lead to unacceptable corrections even if they are localized. We do not discuss here what the magnitude of the corrections is.

The $z$-coordinate part of a KK mode $\psi_{(ml)}(z)$ satisfies the Schr\"{o}dinger-like equation~\eqref{eq:Schrodinger} and one can find a general solution with zero eigenvalue $m^2=l^2$ (see appendix~\ref{appendix3}):
\begin{align}\label{zeroeigenvaluesol}
  \psi_{m^2=l^2}(z)=\left[a^3(z)b^d(z)f_{R}(z)\right]^{1/2}\left[C_1+ C_2 \int
  \frac{1}{a^3(z)b^d(z)f_{R}(z)}\mathrm{d}z\right]\,,
\end{align}
where $C_1$ and $C_2$ are integration constants. There exists the possibility of the normalization of KK modes with this general solution. Moreover, the KK modes of the graviton propagating in the bulk are subjected to boundary conditions~\cite{Gherghetta:2010}. For our case, the vanishing of the variation of the action of KK gravitons at the boundary $\partial\Sigma$ of the bulk along the fifth dimension $z$ leads to the boundary condition
\begin{align}\label{boundarycondition}
  \delta h^{\mu\nu}\partial_z h_{\mu\nu}\big|_{\partial\Sigma}=0\,.
\end{align}
Note that $h_{\mu\nu}$ is assumed to vanish at the four-dimensional boundary $x^{\rho}\rightarrow \pm \infty$ and the extra $d$-dimensional boundary $w^i\rightarrow \pm \infty$, and corresponding boundary terms are automatically zero. Furthermore, the Schr\"{o}dinger-like equation~\eqref{eq:Schrodinger} is a Sturm-Liouville equation and $\psi_{ml}(z)$ satisfies the following orthonormal condition
\begin{align}\label{orthonormalcon}
  \int\mathrm{d}z\;\psi_{(ml)}(z)\psi_{(m^{\prime}l^{\prime})}(z)=\delta_{l\,l^{\prime}}\delta_{m\;m^{\prime}}\,.
\end{align}
Accordingly, the boundary condition~\eqref{boundarycondition} can again be satisfied by imposing either
\begin{align}\label{DirichletNeumann}
  \text{Dirichlet condition:}& \quad \left[a^3(z)b^d(z)f_{R}(z)\right]^{-1/2}\psi_{(ml)}\left(z\right)\bigg|_{\partial\Sigma}=0\,,\\
  \text{or Neumann condition:}& \quad \partial_z\left\{\left[a^3(z)b^d(z)f_{R}(z)\right]^{-1/2}\psi_{(ml)}\left(z\right)\right\}\bigg|_{\partial\Sigma}=0\,.
\end{align}
The solutions of the Schr\"{o}dinger-like equation~\eqref{eq:Schrodinger} need to satisfy these boundary conditions.

The Dirichlet condition may allow the modes with the second particular solution
\begin{align}\label{secondps}
  \psi_{m^2=l^2}(z)=C_2\left[a^3(z)b^d(z)f_{R}(z)\right]^{1/2} \int \frac{1}{a^3(z)b^d(z)f_{R}(z)}\mathrm{d}z\,.
\end{align}
The Neumann condition leads to the modes with the general solution~\eqref{zeroeigenvaluesol}. Particularly, the first particular solution
\begin{align}\label{firstps}
  \psi_{m^2=l^2}(z)=C_1\left[a^3(z)b^d(z)f_{R}(z)\right]^{1/2}
\end{align}
always meets the Neumann condition for arbitrary warp factors and the form of $f(R)$. For an AdS bulk or an asymptotically AdS one, the warp factors may be exponentially divergent at the fifth-dimensional boundary $z\rightarrow \pm \infty$ and this particular solution may still not be normalizable. Thus, only the (asymptotically) behavior of the bulk is not be sufficient enough. As long as the particular solution~\eqref{firstps} is normalizable, one can get a four-dimensional effective theory. It is worth noting that there is the graviton zero mode $\psi_{(00)}(z)$ with the first particular solution~\eqref{firstps}
\begin{equation}\label{zeromode}
  \psi_{(00)}(z)=C_1\left[a^3(z)b^d(z)f_{R}(z)\right]^{1/2}\,.
\end{equation}
Consequently, the mode $\psi_{(00)}(z)$ should satisfy the normalization condition~\eqref{normalizationc}. In this case, the four-dimensional effective Planck mass $M_{\textrm{Pl}}$ is given by
\begin{align}\label{Plz}
  M_{\textrm{Pl}}^2=M_{(D)}^{D-2}V\int\mathrm{d}z\;a^3(z)b^d(z)f_R(z)\,.
\end{align}
Essentially, one does not rule out the possibility that the general solution~\eqref{zeroeigenvaluesol} or the second particular solution~\eqref{secondps} met the boundary condition~\eqref{boundarycondition} is normalizable for some warp factors and the specific form of $f(R)$. Furthermore, additional conditions, such as several kinds of instabilities of $f(R)$ theory, impose restrictions on these solutions of the graviton zero mode. In the following, we investigate brane solutions in the first particular solution~\eqref{zeromode} of the graviton zero mode.

\section{Brane Solutions in Six-dimensional Spacetime}
\label{sec:6dmodel}

In this section, we will seek brane solutions satisfying the above restrictive conditions. As an example, we consider a background real scalar field within the context of six-dimensional $f(R)$ gravity.
Different from the line element with two compact extra dimensions~\cite{Bronnikov:2017bpr,Bronnikov:2020bpr}, we consider the following one
\begin{equation}\label{6linee}
\mathrm{d}s^2=a^2(y)\eta_{\mu\nu}\mathrm{d}x^\mu\mathrm{d}x^\nu+\mathrm{d}y^2+
L^2b^2(y)\mathrm{d}\theta^2\,,
\end{equation}
where $\theta \in [0, 2\pi)$ is a compact dimension. The background scalar field $\phi$ is merely a function of $y$ for a static flat brane. From Eq.~\eqref{eq:eq1MN}, we obtain the following equations in six dimensions
\begin{subequations}\label{eq:6deq1MN}
\begin{align}
&(\mu,\nu): & \frac{1}{2}f+\left(3\frac{a'^{2}}{a^{2}}+\frac{a''}{a}+\frac{a'b'}{ab}\right)f_{R}&-\left(3\frac{a'}{a}+
\frac{b'}{b}\right)f_{R}'-f_{R}''=\kappa_6^{2}\left[\frac{1}{2}\phi^{\prime2}+V(\phi)\right],\label{eq:6deq1compuv}\\
&(y,y): & \frac{1}{2}f+\left(4 \frac{a''}{a}+\frac{b''}{b}\right)f_{R}&-\left(4 \frac{a'}{a}+\frac{b'}{b}\right)f_{R}'
=-\kappa_6^{2}\left[\frac{1}{2}\phi^{\prime2}-V(\phi)\right]\,,\label{eq:6deq1comp55}\\
&(\theta,\theta): & \frac{1}{2}f+\left(\frac{b''}{b}+4 \frac{a'b'}{ab}\right)f_{R}&-4 \frac{a'}{a} f_{R}^{\prime}- f_{R}^{\prime \prime}=\kappa_6^{2}\left[\frac{1}{2}\phi^{\prime2}+V(\phi)\right]\,.\label{eq:6deq1comp66}
\end{align}
\end{subequations}
The equation of motion for the background scalar field is
\begin{equation}\label{eq:6deq2phi}
\phi''+\left(4 \frac{a'}{a}+\frac{b'}{b}\right)\phi'-\frac{\partial V}{\partial \phi}=0\,.
\end{equation}
The combination of Eq.~\eqref{eq:6deq1MN} yields three transformed equations
\begin{subequations}\label{eq:6deq1MNtrans}
\begin{align}
  \left(3\frac{a'^2}{a^2}-3\frac{a' b'}{a b}+\frac{a''}{a}-\frac{b''}{b}\right)f_{R}
+\left(\frac{a'}{a}-\frac{b'}{b}\right)f_{R}'&=0\,,  \\
  \left(3\frac{a'^2}{a^2}+\frac{a'b'}{a b}-3\frac{a''}{a}-\frac{b''}{b}\right)f_{R}
+\frac{a'}{a}f_{R}'-f_{R}''&=\kappa_6^2 \phi'^2\,, \\
  4 \left(\frac{a' b'}{a b}-\frac{a''}{a}\right)f_{R}+\frac{b'}{b}f_{R}'-f_{R}''&=\kappa_6^2 \phi'^2 \,.
\end{align}
\end{subequations}
Here, only three of the four equations in~\eqref{eq:6deq2phi} and \eqref{eq:6deq1MNtrans} are independent because the energy momentum tensor is conserved. If the two warp factors satisfy $a(y)=b(y)$, we have only two independent equations.

For the following tensor perturbations
\begin{equation}\label{perturbation6d}
\mathrm{d}s^2=a^2(y)(\eta_{\mu\nu}+h_{\mu\nu})\mathrm{d}x^\mu\mathrm{d}x^\nu+\mathrm{d}y^2+
L^2b^2(y)\mathrm{d}\theta^2\,,
\end{equation}
the main perturbed equation is
\begin{align}\label{eq:perturbedequvend6d}
f_{R}\left[\frac{1}{a^2}\square^{(4)} h_{\mu \nu}+\frac{1}{L^2b^2}\partial_\theta\partial_\theta h_{\mu \nu}+\left(4 \frac{\partial_y a}{a}+\frac{\partial_y b}{b}\right) \partial_y h_{\mu \nu}+ \partial_{y} \partial_{y} h_{\mu \nu}\right]+\partial_y f_{R} \partial_{y} h_{\mu \nu}=0\,.
\end{align}
The relationship between the four-dimensional effective Planck scale $M_{\textrm{Pl}}$ and the six-dimensional fundamental scale $M_{(6)}$ is
\begin{align}\label{Pl6}
  M_{\textrm{Pl}}^2=M_{(6)}^{4}2\pi L\int\mathrm{d}y\;a^2(y)b(y)f_R(y)\,.
\end{align}

So far, for six-dimensional $f(R)$ gravity within brane world scenarios, we have obtained background equations of motion and have analyzed linear stability for tensor perturbations. We will solve analytical solutions in the following subsections with Eqs.~\eqref{eq:6deq2phi} and~\eqref{eq:6deq1MNtrans}.

\subsection{Starobinsky Gravity}

First, we consider the Starobinsky gravity
\begin{equation}\label{starobinsky}
  f(R)=R+\alpha R^2,
\end{equation}
which is the first model of inflation in four dimensions and also gives rise to cosmic acceleration which ends when the term $\alpha R^2$ is smaller than $R$~\cite{Starobinsky:1980s}. Here, we seek thick brane solutions in the six-dimensional Starobinsky gravity.

In general, the superpotential method is not available for higher-order differential equations. In order to avoid solving  higher-order equations directly, we would like to adopt the reconstruction technique~\cite{Higuchi:2014hn,Chakraborty:2016cs,Cui:2018clgz}, which seeks a reasonable action of background fields to satisfy fixed configurations reasonable for brane models. As mentioned above, for simplicity we set the warp factors $a(y)$ and $b(y)$ as
\begin{equation}\label{warpfactor1}
  a(y)=b(y)=\mathrm{sech}^n(k y)\,,
\end{equation}
where $k$ is a parameter with dimension mass. For $ky\rightarrow\pm\infty$, $a(y)=b(y)\rightarrow \mathrm{e}^{-nk|y|}$, which implies that the metric~\eqref{6linee} reduces an AdS one at the boundary. The AdS curvature relates to the parameters $k$ and $n$. The length scale $1/k$ corresponds to the thickness of the thick brane.

\begin{figure}[htb]
  \centering
  \includegraphics[width=2.6in]{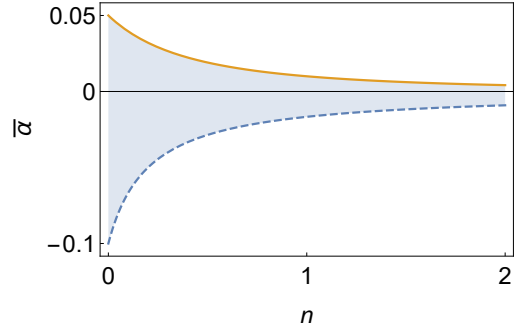}
  \caption{The domain of existence of the solution (\ref{solution1phi}) in the $(\bar{\alpha},n)$ parameter space. The blue dashed line stands for the lower bound and is not an available curve of the parameter. The orange solid line stands for the upper bound and is an available curve of the parameter. The region between the blue dashed line and the orange solid line is an existence region of the solution for $n>0$. We only give a schematic diagram in the domain $n\in (0,2)$. It is worth noting that two bound lines of $\bar{\alpha}$ will approach to zero for $n\rightarrow \infty$. }\label{fig:pspace1}
\end{figure}

\begin{figure}[htb]
\centering
\subfigure[\ The scalar field $\bar{\phi}(\bar{y})$. \label{subfig:phin1}]{
\includegraphics[width=2.6in]{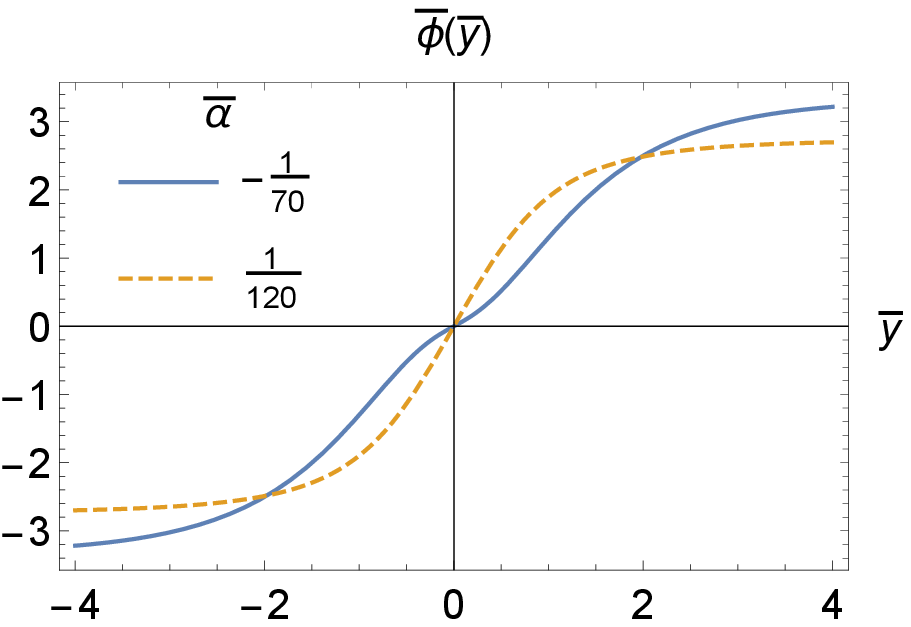}}
\hspace{0.5in}
\subfigure[\ The scalar potential $\bar{V}(\bar{\phi})$. \label{subfig:phin2}]{
\includegraphics[width=2.6in]{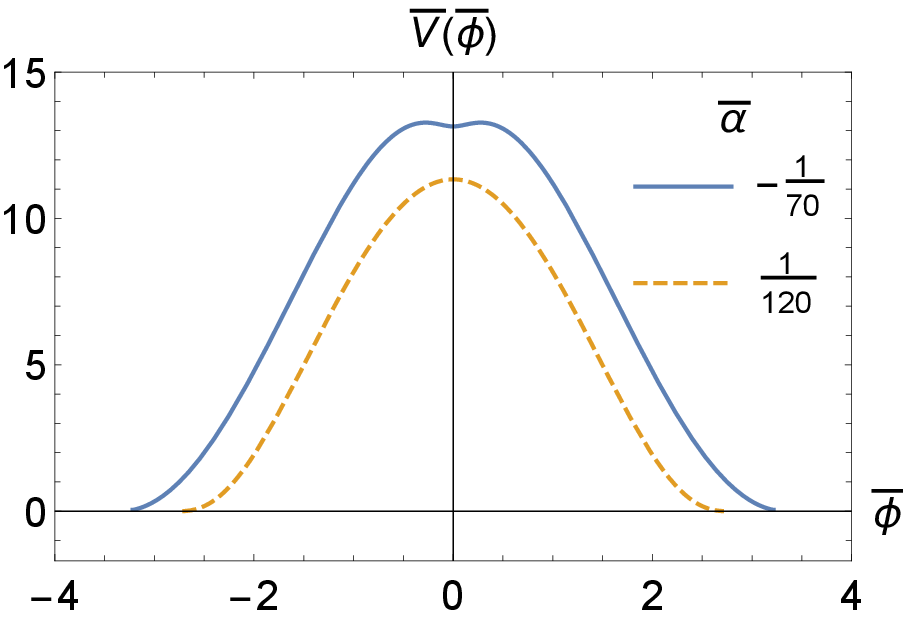}}
\caption{Plots of the dimensionless scalar field $\bar{\phi}(\bar{y})$ and the corresponding dimensionless scalar potential $\bar{V}(\bar{\phi})$ for the brane world solution \eqref{solution1}. The parameter $n$ is set to $1$.}\label{fig:phin}
\end{figure}

From three independent equations in Eqs.~\eqref{eq:6deq2phi} and~\eqref{eq:6deq1MNtrans}, we obtain the following analytic solution
\begin{subequations}\label{solution1}
\begin{align}
  \phi(y)~~=&\pm\frac{2 \sqrt{n}\sqrt{q}}{\kappa_6 } \left\{\mathrm{i} \left[\mathrm{E}\left(\mathrm{i} k y,\frac{p}{q}\right)-\mathrm{F}\left(\mathrm{i} k y,\frac{p}{q}\right)\right]+ \sqrt{1+\frac{p }{q}\sinh ^2(k y)}\tanh (k y)\right\}\,, \label{solution1phi} \\
  V(\phi(y))=&\frac{k^2}{\kappa_6^2}2 n\left[(5 n+1) p\,\mathrm{sech}^2(k y)+(5 n+2) \xi\, \mathrm{sech}^4(k y)\right]\,,\label{solution1Vphi}
\end{align}
\end{subequations}
with three dimensionless parameters
\begin{align}
  p=1-10 k^2\alpha (n+1) (3 n+2)\,, \quad
  q=1+10 k^2\alpha (5 n+1)\,, \quad
  \xi=5 k^2\alpha (n+3) (3 n+1)\,. \nonumber
\end{align}
Here $\mathrm{F}$ and $\mathrm{E}$ are the first and second elliptic integrals, respectively. The signs $\pm$ in Eq.~\eqref{solution1phi} stand for two solutions with opposite values. That the solution \eqref{solution1phi} is real requires
\begin{align}\label{conditionofsolution}
  -\frac{1}{10(5n+1)}<k^2\alpha \leqslant\frac{1}{10(n+1)(3 n+2)}.
\end{align}
For the sake of clarity, we introduce the following dimensionless quantities
\begin{align}
  \bar{y}=k y, \quad \bar{\alpha}=k^2\alpha, \quad \bar{\phi}=\kappa_6\phi, \quad \bar{V}=\kappa_6^2V/k^2.
\end{align}
The parameter space allowing brane solutions is shown in Fig.~\ref{fig:pspace1}.
Moreover, the scalar field $\phi(y)$ and the scalar potential $V(\phi)$ in \eqref{solution1} are shown in Fig.~\ref{fig:phin}.
It is shown that an AdS domain wall solution is characterized by a function $\bar{\phi}(\bar{y})$ that approaches to the two different vacua of the potential as $\bar{y}\rightarrow\pm\infty$. This solution also indicates a length scale corresponding to the thickness of the brane.

\begin{figure}[htb]
\centering
\subfigure[\ The scalar field $\bar{\phi}(\bar{y})$.  \label{subfig:phic1}]{
\includegraphics[width=2.6in]{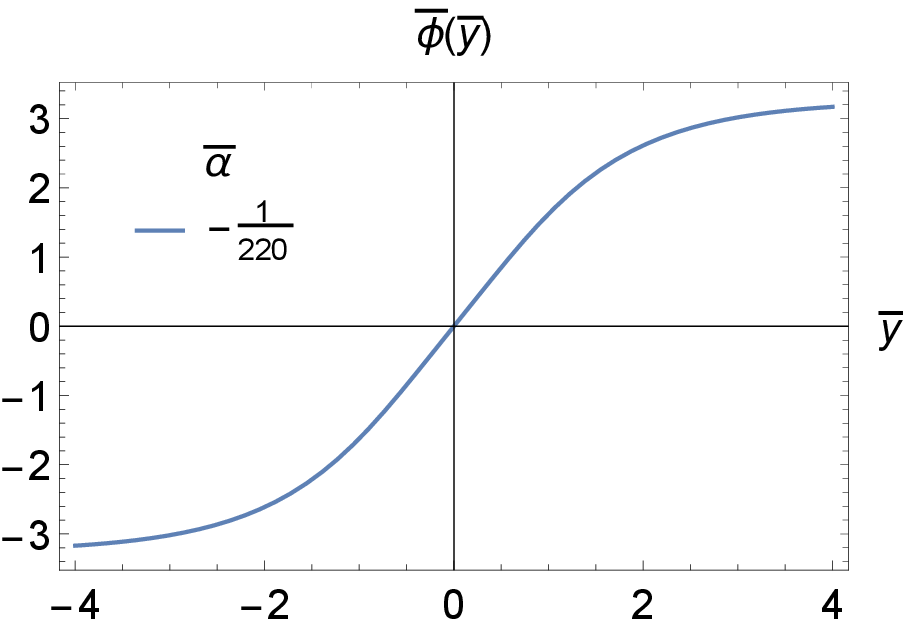}}
\hspace{0.5in}
\subfigure[\ The scalar potential $\bar{V}(\bar{\phi})$.  \label{subfig:Vc1}]{
\includegraphics[width=2.6in]{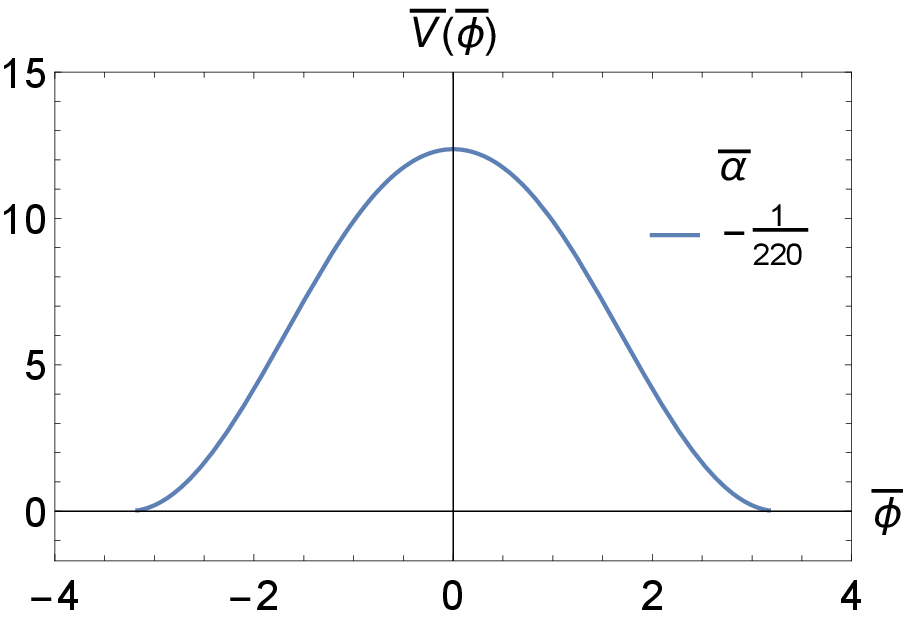}}
\caption{Plots of the dimensionless scalar field $\bar{\phi}(\bar{y})$ at the critical value $\bar{\alpha}_\text{c}$ and the corresponding dimensionless scalar potential $\bar{V}(\bar{\phi})$. The parameter $n$ is set to $1$.}\label{fig:phiVc1}
\end{figure}

From Fig.~\ref{subfig:phin1}, the scalar field~\eqref{solution1phi} is a double kink for small values of $\bar{\alpha}$ but is a single kink for large values of $\bar{\alpha}$. There is a critical value at which the double kink becomes a single one. We can seek the critical value by requiring the third-order derivative of the dimensionless scalar field~\eqref{solution1phi} with respect to $\bar{y}$ to satisfy the following condition
\begin{equation}
  \frac{\mathrm{d}^3\bar{\phi}(\bar{y})}{\mathrm{d}\bar{y}^3}\bigg|_{\bar{y}=0}=0\,.
\end{equation}
The critical value of $\bar{\alpha}$ is given by
\begin{align}\label{criticalfv}
  \bar{\alpha}_{\text{c}}=-\frac{1}{30 n (n+5)+40}\,,
\end{align}
at which
\begin{align}
 \bar{\phi}'(0)=\pm {2 \sqrt{n}}\sqrt{\frac{(n+3) (3 n+1)}{3 n (n+5)+4}}\,.
\end{align}
This means that when $\bar{\alpha}$ exceeds the critical value $\bar{\alpha}_\text{c}$, the scalar field~\eqref{solution1phi} becomes a single kink. We exhibit the critical solution of the scalar field and the critical scalar potential in Fig.~\ref{fig:phiVc1}.

\begin{figure}[htb]
\centering
\subfigure[\ The scalar field $\bar{\phi}(\bar{y})$. \label{subfig:phic2}]{
\includegraphics[width=2.6in]{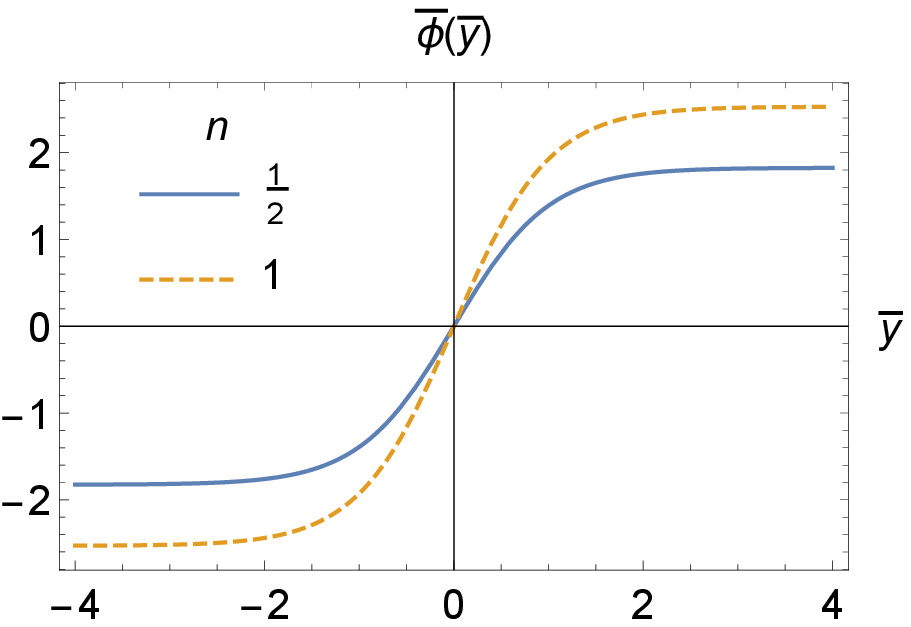}}
\hspace{0.5in}
\subfigure[\ The scalar potential $\bar{V}(\bar{\phi})$. \label{subfig:Vc2}]{
\includegraphics[width=2.6in]{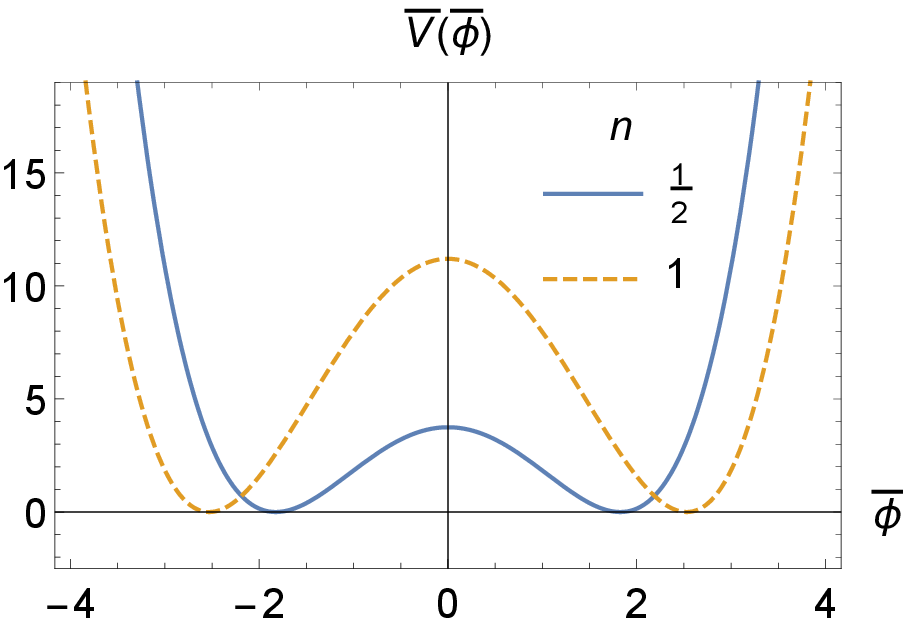}}
\caption{Plots of the dimensionless scalar field $\bar{\phi}(\bar{y})$ and the corresponding dimensionless scalar potential $\bar{V}(\bar{\phi})$ in Eq.~\eqref{solution1c}. }\label{fig:phiVc2}
\end{figure}

For the special value of $\bar{\alpha}$:
\begin{align}
  \bar{\alpha}=\frac{1}{10(n+1)(3 n+2)},
\end{align}
we obtain a concise solution
\begin{subequations}\label{solution1c}
\begin{align}
  \phi(y)=&\pm  v\tanh (k y)\,, \label{solution1cphi} \\
  V(\phi)=&\frac{(5 n+2)}{4}\frac{k^2}{v^2}\left(\phi^2-v^2\right)^2\,, \label{solution1cVphi}
\end{align}
\end{subequations}
where
\begin{align}
  v=\frac{1}{\kappa_6}\sqrt{\frac{4 n (n+3) (3 n+1)}{(n+1) (3 n+2)}}. \nonumber
\end{align}
We depict the kink solution and the scalar potential in Fig.~\ref{fig:phiVc2}.

Next we explore the distribution of the energy density $\rho$ along the extra dimension $y$:
\begin{align}\label{edensity}
   \rho= T_{MN}U^M U^N ,  
\end{align}
where $U^M=(1/a(y),0,0,0,0,0)$.
It is worth mentioning that the vacuum energy density has to been deducted from the total energy density. We introduce the dimensionless energy density $\bar{\rho}=\rho\kappa_6^2/k^2$. It is shown that there is a critical value of $\bar{\alpha}$ blow which the energy density will be split. Such critical value of $\bar{\alpha}$ is
\begin{equation}\label{cvofrho}
  \bar{\alpha}_0=-\frac{5 n+2}{10 [n (31 n+35)+8]}\,,
\end{equation}
which is solved from the following condition
\begin{equation}\label{condition}
  \frac{\mathrm{d}^2\bar{\rho}(\bar{y})}{\mathrm{d}\bar{y}^2}\bigg|_{\bar{y}=0}=0\,.
\end{equation}

\begin{figure}[htb]
\centering
\subfigure[\ $\bar{\alpha}=-\frac{1}{10(5n+1)+1}$. \label{fig:energydensitysplit}]{
\includegraphics[width=2.5in]{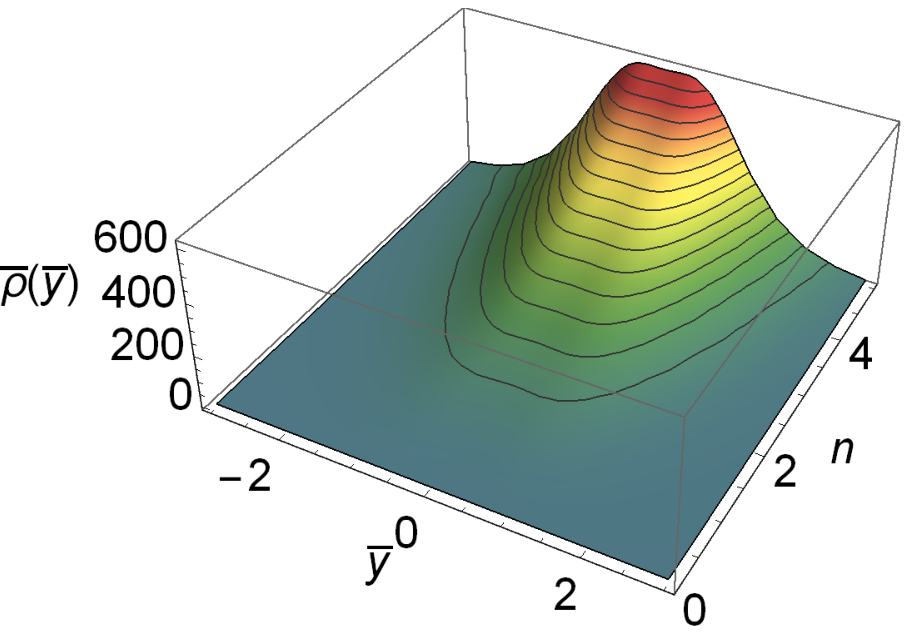}}
\hspace{0.5in}
\subfigure[\ $\bar{\alpha}=\frac{1}{10(n+1)(3n+2)}$. \label{fig:energydensitynonsplit}]{
\includegraphics[width=2.5in]{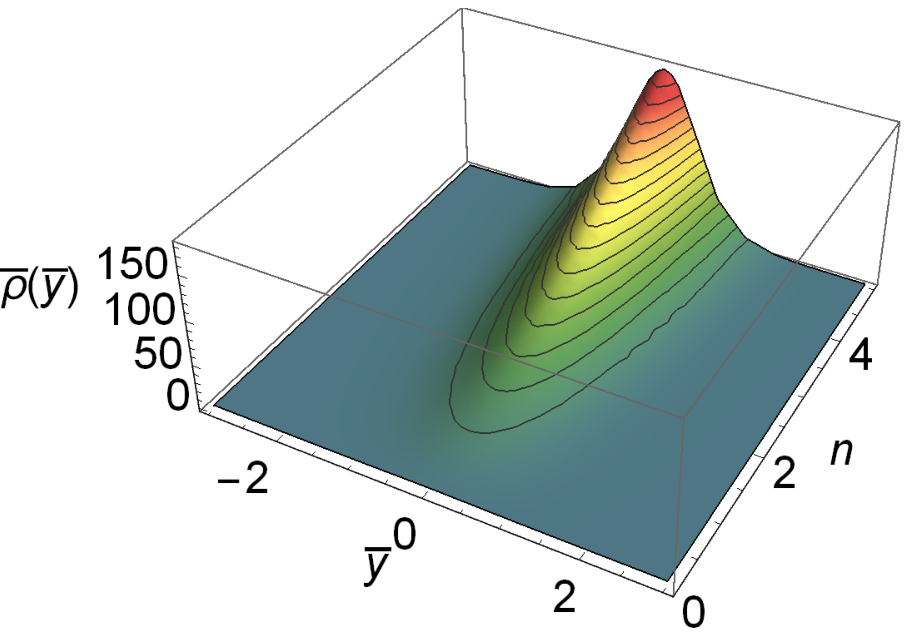}}
\hspace{0.5in}
\subfigure[\ $\bar{\alpha}=\bar{\alpha}_0=-\frac{5 n+2}{10 [n (31 n+35)+8]}$. \label{fig:energydensityc}]{
\includegraphics[width=2.5in]{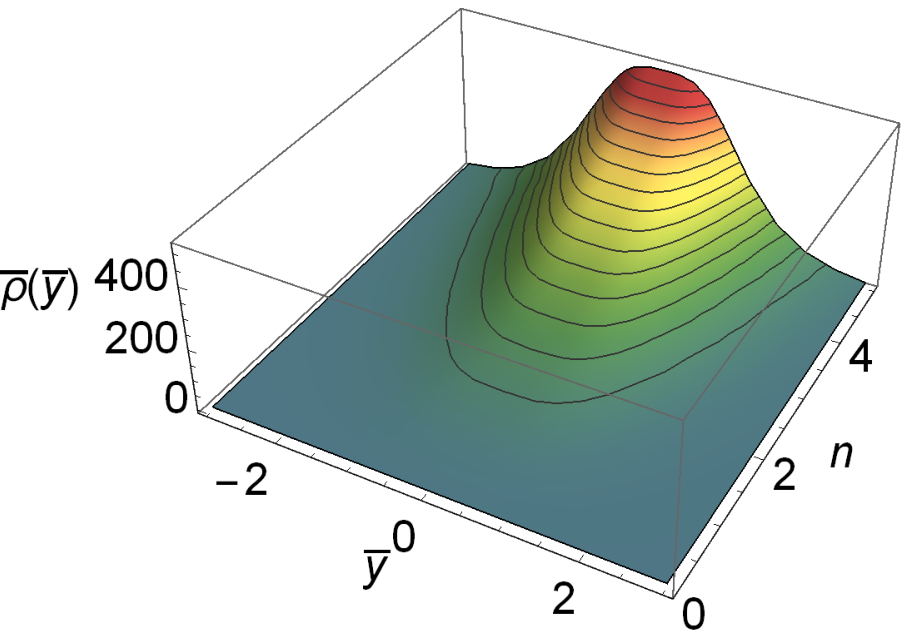}}
\caption{Plots of the dimensionless energy density $\bar{\rho}(\bar{y})$ for the solution \eqref{solution1} regarding three values of $\bar{\alpha}$. The energy density in Fig.~\ref{fig:energydensitysplit} exhibits split behavior. A non-split energy density is shown in Fig.~\ref{fig:energydensitynonsplit}. The energy density at the critical value $\bar{\alpha}_0$ is shown in Fig.~\ref{fig:energydensityc}. }\label{fig:energydensity}
\end{figure}
\begin{figure}
\centering
\subfigure[\ $n=1/10$.]{
\includegraphics[width=2.5in]{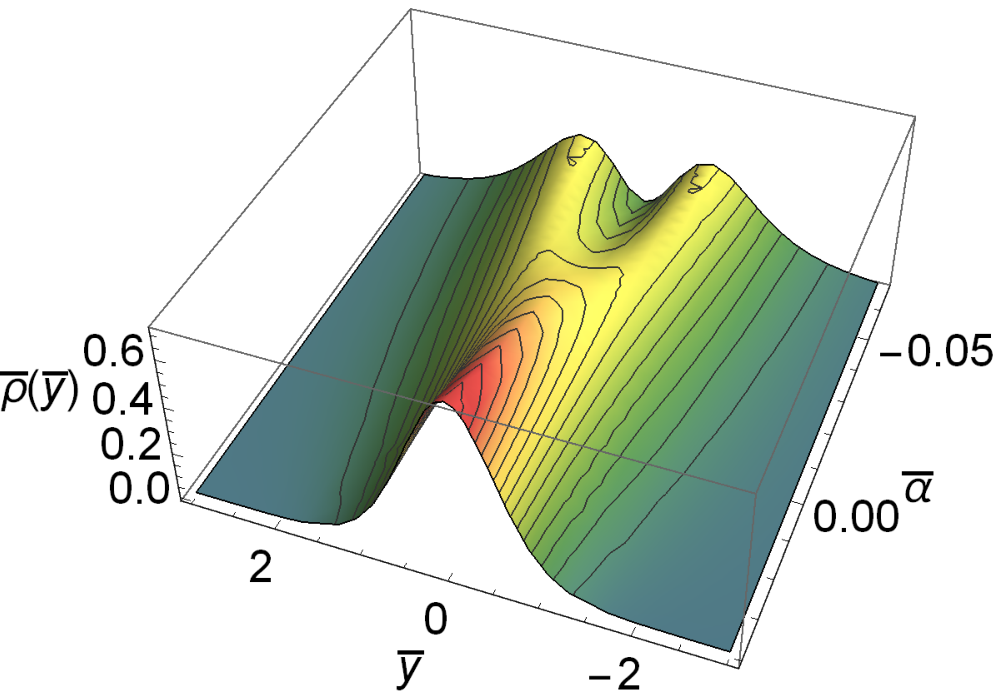}}
\hspace{0.5in}
\subfigure[\ $n=5$.]{
\includegraphics[width=2.5in]{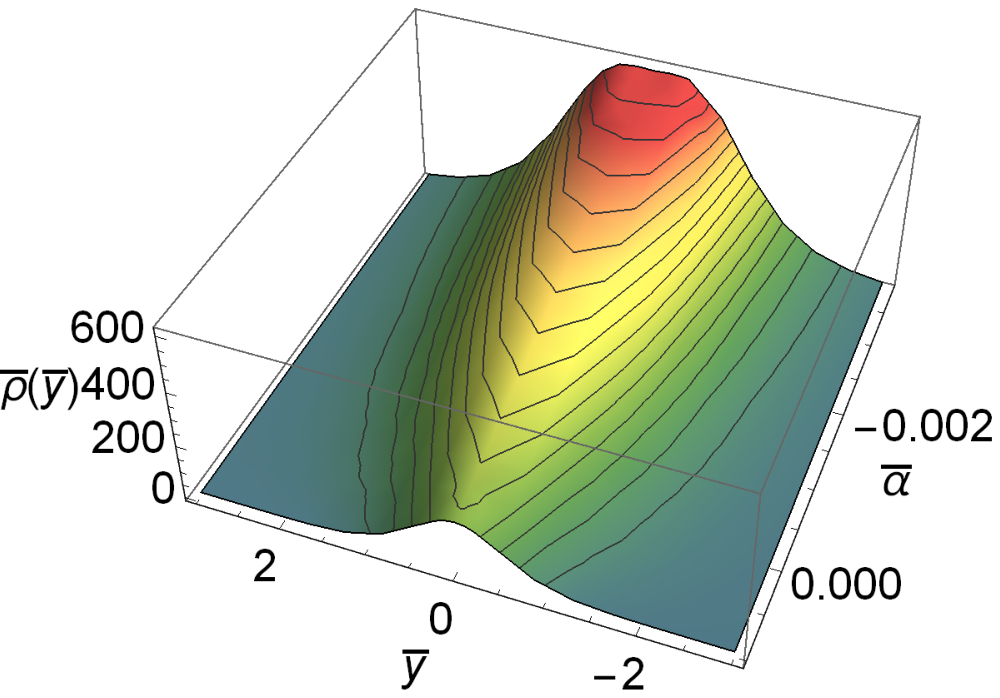}}
\caption{Plots of the dimensionless energy density $\bar{\rho}(\bar{y})$ for the solution \eqref{solution1} regarding two values of $n$.}\label{fig:edpsapcecn}
\end{figure}
For the above solution~\eqref{solution1}, we plot the energy density of the thick brane in Fig.~\ref{fig:energydensity}. It is shown that, for various values of $\bar{\alpha}$, the maximum of the energy density increases with $n$. On the other hand, the brane splits only when $\bar{\alpha}$ is less than the critical value $\bar{\alpha}_0$ (see Figs.~\ref{fig:energydensitysplit} and \ref{fig:energydensitynonsplit}). For a fixed $n$, as shown in Fig.~\ref{fig:edpsapcecn}, the thickness of the brane decreases with increasing $\bar{\alpha}$, which can also be seen from Figs.~\ref{fig:energydensitysplit} and \ref{fig:energydensitynonsplit}, and the brane gradually splits as the decrease of $\bar{\alpha}$.


\subsection{\texorpdfstring{$f(R)=R+\alpha R^3$}{f(R)=R+\alpha R^3} Gravity}

\begin{figure}[htb]
  \centering
  \includegraphics[width=2.6in]{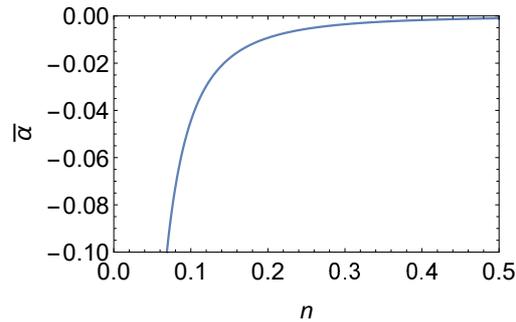}
  \caption{The line of existence of the solution in the $(\bar{\alpha},n)$ parameter space in $f(R)=R+\alpha R^3$ gravity. The blue solid curve stands for the parameter range of our solution for $n>0$. We only give a schematic diagram in the domain $n\in (0,0.5)$. It is worth noting that the blue line of $\bar{\alpha}$ approaches to zero from an infinite negative value with increasing $n$.}\label{fig:pspace2}
\end{figure}

\begin{figure}[htb]
\centering
\subfigure[\ The scalar field $\bar{\phi}(\bar{y})$. \label{subfig:phin21}]{
\includegraphics[width=2.6in]{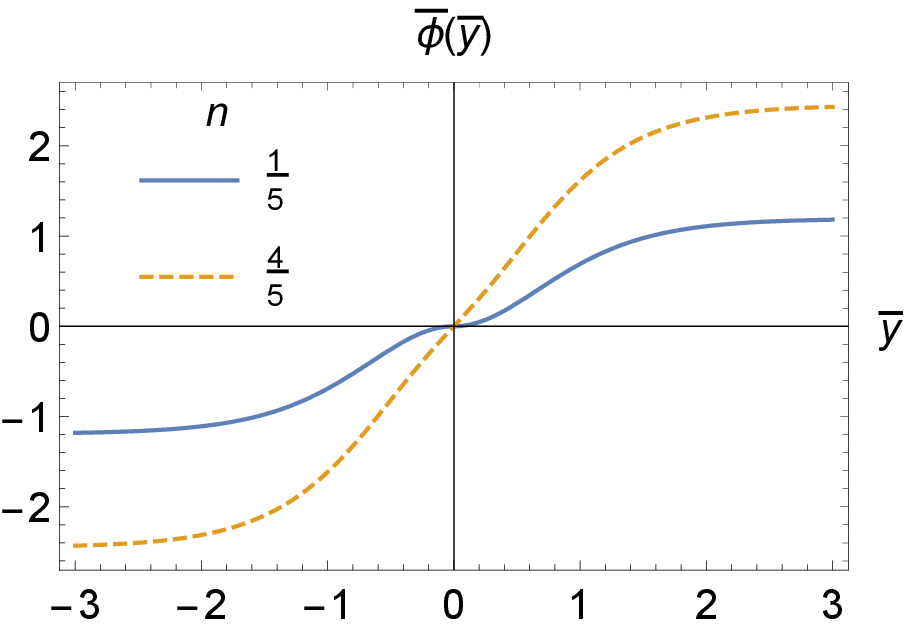}}
\hspace{0.5in}
\subfigure[\ The scalar potential $\bar{V}(\bar{\phi})$. \label{subfig:phin22}]{
\includegraphics[width=2.6in]{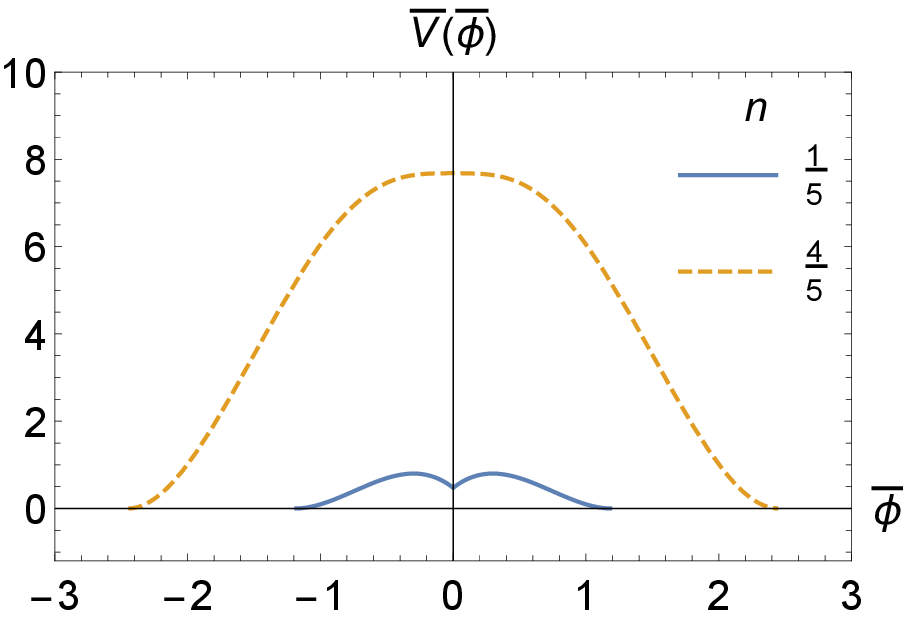}}
\caption{Plots of the dimensionless scalar field $\bar{\phi}(\bar{y})$ and the corresponding dimensionless scalar potential $\bar{V}(\bar{\phi})$ in~\eqref{solution2}. }\label{fig:phin2}
\end{figure}

In this subsection, we consider the following $f(R)$ model
\begin{equation}\label{cubic}
  f(R)=R+\alpha R^3\,,
\end{equation}
where $\alpha$ has dimension $[\text{Length}]^{-4}$. We also consider the warp factors given in Eq.~\eqref{warpfactor1}. In this model, we take the dimensionless parameter
\begin{align}
  \bar{\alpha}=\alpha k^4=-\frac{1}{900n^2(5n+2)}
\end{align}
to solve the equations of motion~\eqref{eq:6deq2phi} and \eqref{eq:6deq1MNtrans}. We show the parameter space of the solution in Fig.~\ref{fig:pspace2}.
It should be noted that solutions in other parameter space out of the blue line in Fig.~\ref{fig:pspace2} may exist. Our solution is given by
\begin{subequations}\label{solution2}
\begin{align}
  \phi(y)=& \pm\frac{1}{\kappa_6}\sqrt{\frac{1}{15 (5 n+2)}} \left[c(y) d(y) \mathrm{sech}^2(k y)+(5 n-1) \mathrm{arctanh}\left(\frac{c(y)}{d(y)}\right)\right]  \,, \\
  V(\phi(y))=& \frac{2 k^2 (3 n+1)  }{9 \kappa_6^2 (5 n+2)} \text{sech}^6(k y) \,\big[3 (5 n+1) (5 n+2) \cosh (2 k y)-25 n-9\big]\,,
\end{align}
\end{subequations}
where the two functions $c(y)$ and $d(y)$ are
\begin{align}
  c(y)=\sqrt{5(3 n+1)} \sinh (k y)\,, \quad d(y)=\sqrt{4 (5 n+1) \sinh^2(k y)+5 n-1}\,.
\end{align}
This solution requires $n\geqslant1/5$. We plot the scalar field $\phi$ and the potential $V(\phi)$ in Fig.~\ref{fig:phin2}. For a small $n$, $\phi$ is a double kink solution.

It is interesting that a simplified solution is obtained if $n=1/5$:
\begin{subequations}\label{solution2c}
\begin{align}
  \phi(y)=& 
  \pm\frac{8}{\bar{\kappa}_6}\mathrm{sgn}(ky) \tanh^2(ky)\,, \\
  V(\phi(y))=&\frac{32 k^2}{3 \bar{\kappa}_6^2} \left[9 \cosh (2 ky)-7\right] \text{sech}^6(ky)\,,
\end{align}
\end{subequations}
where a newly defined coupling constant $\bar{\kappa}_6=3\sqrt{5}\kappa_6$ is adopted. In this special case, we can obtain an analytical expression for the scalar potential $V(\phi)$:
\begin{align}\label{solution2cVphi}
  V(\phi)=& \frac{k^2}{3 \bar{\kappa}_6^2} \left[1+\bar{\kappa}_6\mathrm{sgn}(\phi) \phi \right]
  \left[8-\bar{\kappa}_6\mathrm{sgn}(\phi) \phi \right]^2\,.
\end{align}

\begin{figure}[htb]
\centering
\subfigure[\ The energy density $\bar{\rho}(\bar{y})$.]{
\includegraphics[width=2.6in]{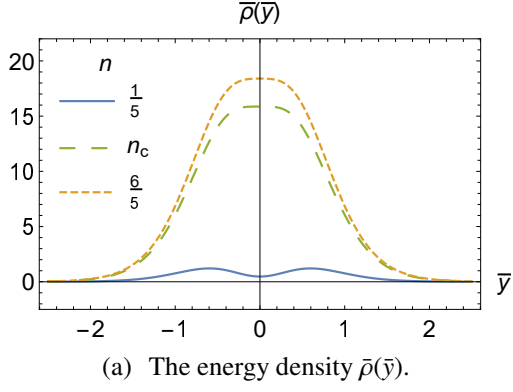}}
\hspace{0.5in}
\subfigure[\ The energy density $\bar{\rho}(\bar{y})$.]{
\includegraphics[width=2.6in]{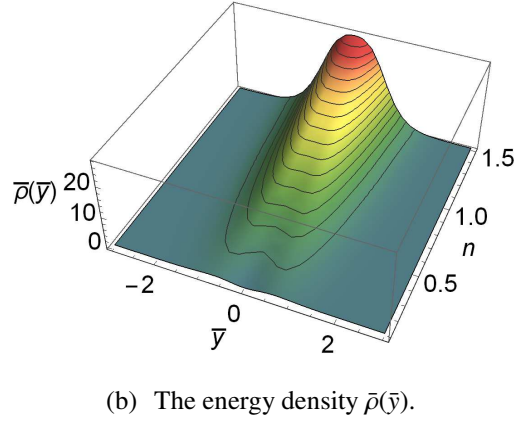}}
\caption{Plots of the dimensionless energy density $\bar{\rho}(\bar{y})$ in Eq.~\eqref{solution2}. The blue solid line and the orange dashed line represent the split and non-split energy densities, respectively. The green long dashed line represents the critical line of energy density split. }\label{fig:edensity2}
\end{figure}

According to Eq.~\eqref{edensity}, we plot the energy density in Fig.~\ref{fig:edensity2}.
It is shown that the brane is split for a small $n$. There is a critical value
\begin{align*}
  n_{\mathrm{c}}=(\sqrt{65}+3)/10 \simeq 1.106\,,
\end{align*}
which satisfies the condition~\eqref{condition}. The brane is not split for $n>n_{\mathrm{c}}$. Besides, the maximum of the energy density increases with $n$.

\section{Effective Potentials}
\label{sec:epotential}

\begin{figure}[htb]
\centering
\subfigure[\ $f(R)=R$. \label{subfig:W(z)GR}]{
\includegraphics[width=2.6in]{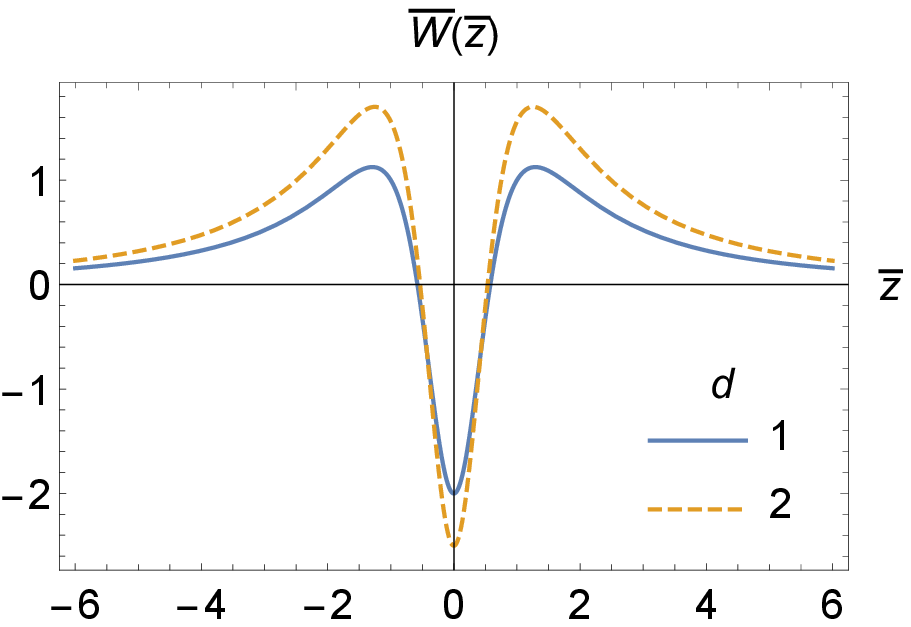}}
\hspace{0.5in}
\subfigure[\ $f(R)=R+\alpha R^2$.  \label{subfig:W(z)R2}]{
\includegraphics[width=2.6in]{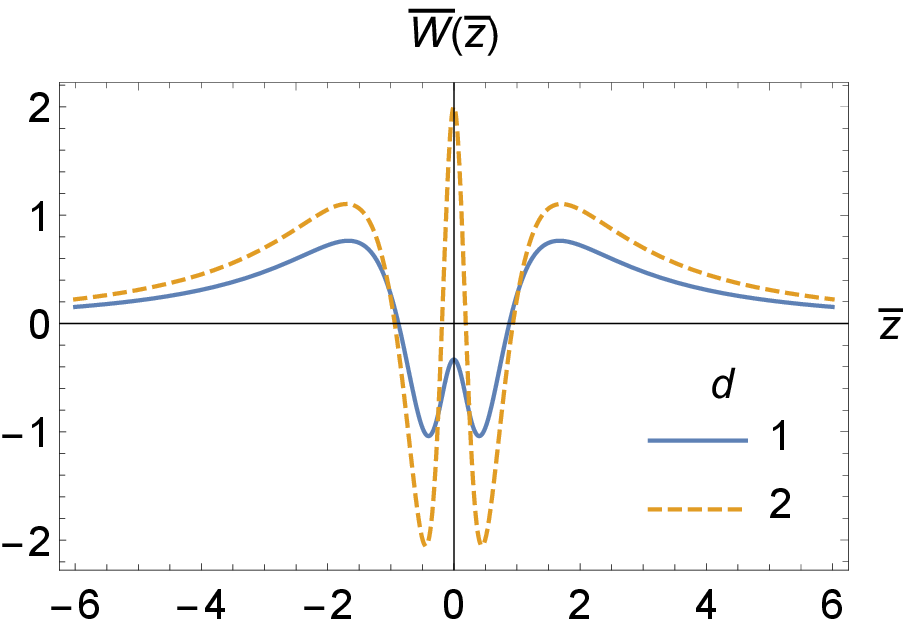}}
\hspace{0.5in}
\subfigure[\ $f(R)=R+\alpha R^3$.  \label{subfig:W(z)R3}]{
\includegraphics[width=2.6in]{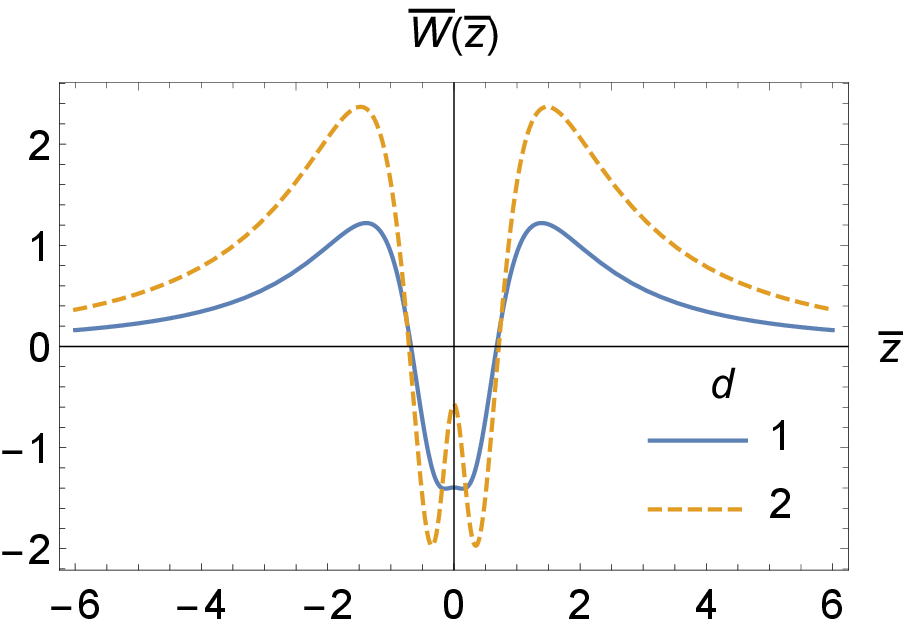}}
\hspace{0.5in}
\subfigure[\ $f(R)=R+\alpha R^2$. \label{subfig:zeromode}]{
\includegraphics[width=2.6in]{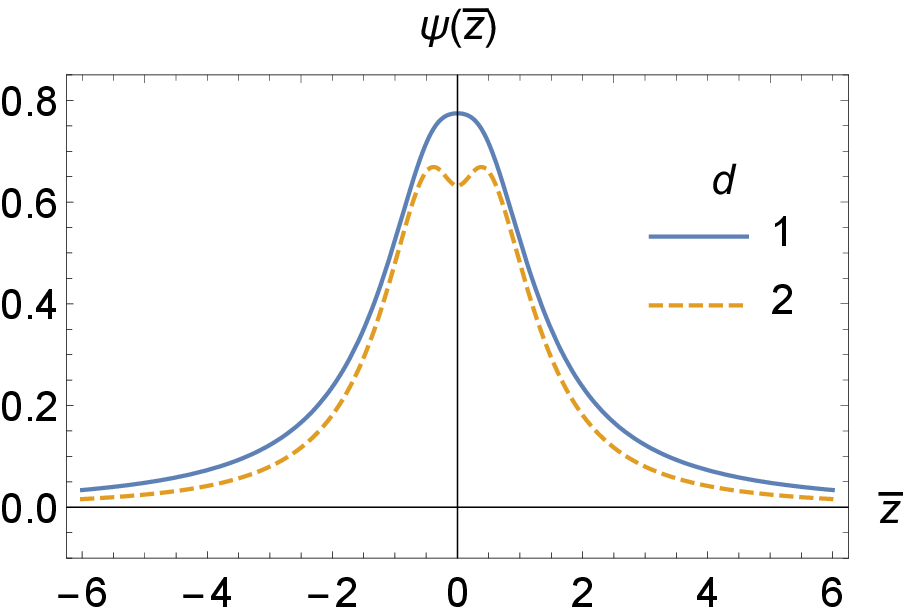}}
\caption{Plots of the dimensionless effective potential $\bar{W}(\bar{z})$ in the Schr\"{o}dinger-like equation~\eqref{eq:Schrodinger} for various $f(R)$ models and the zero mode for $f(R)=R+\alpha R^2$, where $\bar{z}=kz$ and $\bar{W}(\bar{z})=W(\bar{z})/k^2$.  In Figs.~\ref{subfig:W(z)R2} and~\ref{subfig:zeromode}, we adopt $\bar{\alpha}=k^2\alpha=-1/20$. In Fig.~\ref{subfig:W(z)R3}, we adopt $\bar{\alpha}=k^4\alpha=-0.0025$. All the effective potentials approach to zero at $\bar{z}\rightarrow \pm \infty $.}\label{fig:W(z)}
\end{figure}

The above six-dimensional thick brane solutions can be generalized to a higher-dimensional bulk. To explore the effective gravity on the brane in such a background, we discuss the effective potential~\eqref{epotential} for the above solved $f(R)$ models in higher dimensions with the warp factors~\eqref{warpfactor1}. Here we only consider the case $n=1$. For these two solved $f(R)$ models, the graviton zero mode~\eqref{zeromode} satisfies the normalization condition~\eqref{normalizationc} when $n>0$ for the warp factors~\eqref{warpfactor1}. As an example, we show a normalized graviton zero mode in Fig.~\ref{subfig:zeromode}. The difference of the effective potential~\eqref{epotential} between GR and $f(R)$ gravity is demonstrated in Fig.~\ref{fig:W(z)}.

From Fig.~\ref{fig:W(z)}, one can find that the effective potential in Fig.~\ref{subfig:W(z)R2} has a double well due to the curvature scalar correction concerning GR. Therefore, a correction term for GR alters the effective potential for the graviton KK modes. This implies that the graviton mass spectrum will be different for various gravity theories. Besides, the depth of the effective potential is increased with the increase of the dimensions $d$ of the extra space. Therefore, the KK modes of the graviton may partially reflect the configuration of extra dimensions.


\section{Conclusions}
\label{sec:conclusions}

In summary, we investigated $f(R)$ gravity with extra spatial dimensions in brane world scenarios. It was shown that the spacetime configuration is perturbatively stable. Further, we found two sets of thick brane solutions in a six-dimensional spacetime. We showed that the distribution of the energy density of the background scalar field concentrates on the vicinity of the thick brane, which implies that there is no divergence of curvature. Moreover, it was shown that the graviton zero mode is localized on the brane embedded in higher dimensions. This implies that the four-dimensional Newtonian potential can be recovered on the brane.


In this paper, we considered an extra $d$-dimensional Euclidean space $\mathcal{E}_d$ besides the extra dimension $y$. A general extra $d$-dimensional Riemannian space is still interesting. It is well known that $f(R)$ theory is equivalent to a second-order scalar-tensor theory of gravitation. In our exploration, we investigated branes only in $f(R)$ gravity rather than transform it into a scalar-tensor theory. So far, we considered a special six-dimensional example. Models in different dimensions can lead to novel features of the bulk spacetime. The background metric ansatz~\eqref{linee} or~\eqref{6linee} can also be replaced by other forms which are from some interesting configurations of the bulk. We leave this in future works. In this paper, we considered the brane with four-dimensional Poincar\'{e} symmetry. Naturally, one can explore branes with other symmetries, such as four-dimensional dS or AdS branes. Particularly, an investigation of spherically symmetric branes in the presence of $f(R)$ gravity was made in Refs.~\cite{Borzou:2009bssy,Chakraborty:2015csfr,Chakraborty:2016csfrgb}. Moreover, a complete investigation should include the localization of matter fields on the brane. For some of the related references, one may refer to~\cite{Mitra:2017mps,Zhou:2018zdyl,Liu:2017l}. We will not address this issue in this paper.

\section*{Acknowledgements}

The authors would like to thank Bao-Min Gu for helpful discussions. This work is supported in part by the National Natural Science Foundation of China (Grants No. 11875151 and No. 11522541) and the Fundamental Research Funds for the Central Universities (Grants No. lzujbky-2019-it21, No. lzujbky-2019-ct06, and No. lzujbky-2020-it04). 

\appendix

\section{General Perturbations}
\label{appendix1}

Suppose that a $D$-dimensional spacetime undergoes a small perturbation $\delta g_{MN}$ on a fixed background $g_{MN}$
\begin{align}
  \tilde{g}_{MN}=g_{MN}+\epsilon\left(\delta g_{MN}\right),
\end{align}
where $\epsilon$ is a real parameter to indicate the degree of perturbation. In light of perturbation theory, the inverse of the perturbed metric is
\begin{align}
  \tilde{g}^{MN}=g^{MN}-\epsilon\left(\delta g^{MN}\right)+\epsilon^2\left(\delta g^{ML}\delta g^N_L\right)+\cdots\,,
\end{align}
where $\delta g^{MN}=g^{MP}g^{NQ}\delta g_{PQ}$ and ellipses correspond to the more higher-order perturbations. Quantities determined by metric can be obtained accordingly. For the sake of simplicity, one can omit the real parameter $\epsilon$ without confusing. We list some quantities under the above metric perturbations as follows:
\begin{align}
  \tilde{\Gamma}^L_{MN}=& \Gamma^L_{MN}+\delta\Gamma^L_{MN}+\cdots\,, \\
  \tilde{R}^L_{MKN}=& R^L_{MKN}+ 2\nabla_{[K|}\delta\Gamma^L_{M|N]}+\cdots\,, \\
  \tilde{R}_{MN}= & R_{MN}+ 2\nabla_{[L|}\delta\Gamma^L_{M|N]}+ \cdots\,,
\end{align}
where
\begin{align}
  \delta\Gamma^L_{MN}= \frac{1}{2}g^{LP}\left(\nabla_M\delta g_{NP}+\nabla_N\delta g_{PM}-\nabla_P\delta g_{MN}\right)
\end{align}
and square brackets ``${[M|\;\;|N]}$'' appearing in indices signify antisymmetrization in terms of $M$ and $N$. In the region of weak gravitational field, one can decompose the metric into a small perturbation around a flat spacetime.

\section{Explicit Forms of Perturbations of the Fundamental Quantities}
\label{appendix2}

To linear order, the perturbed forms of the fundamental quantities will be collected in this appendix. Firstly, we keep both the zeroth and linear order of perturbed quantities without considering the condition~\eqref{TTcondition}. Secondly, we give necessary simplified quantities by introducing the condition~\eqref{TTcondition}, which will be used in this paper.

We list perturbed quantities up to the linear order without considering condition~\eqref{TTcondition} in the following. The nonvanishing components of the perturbation of the connection are
\begin{subequations}
\begin{align}
\tilde{\Gamma}_{\mu \nu}^{\lambda}=&\frac{1}{2}\left(\partial_{\mu} h_{\nu}^{\lambda}
+\partial_{\nu} h_{\mu}^{\lambda}-\partial^{\lambda} h_{\mu \nu}\right)\,,\\
\tilde{\Gamma}_{\mu \nu}^{y}=&-a^2\left[\frac{\partial_{y}a}{a}\left(\eta_{\mu \nu}
+h_{\mu \nu}\right)+\frac{1}{2}\partial_{y} h_{\mu \nu}\right]\,,\\
\tilde{\Gamma}_{\mu \nu}^{i}=&-\frac{1}{2} \frac{a^{2}}{b^2} \partial^{i} h_{\mu \nu}\,,\\
\tilde{\Gamma}_{\mu y}^{\lambda}=&\tilde{\Gamma}_{y\mu}^{\lambda}=
\frac{\partial_{y}a}{a} \delta_{\mu}^{\lambda}+\frac{1}{2} \partial_{y} h_{\mu}^{\lambda}\,,\\
\tilde{\Gamma}_{\mu i}^{\lambda}=&\tilde{\Gamma}_{i\mu}^{\lambda}=
\frac{1}{2} \partial_{i} h_{\mu}^{\lambda}\,,\\
\tilde{\Gamma}_{ij}^{y}=&-b \partial_{y}b\delta_{ij}\,,\\
\tilde{\Gamma}_{jy}^{i}=&\frac{\partial_{y}b}{b}\delta^i_j\,.
\end{align}
\end{subequations}
The nonvanishing components of the perturbation of the Ricci tensor read as
\begin{subequations}
\begin{align}
\tilde{R}_{\mu \nu}=&-a^2\left[\left(3\frac{(\partial_{y}a)^2}{a^2}+\frac{\partial_{y}\partial_{y}a}{a}+d\frac{\partial_{y}a\partial_{y}b}{ab}\right) (\eta_{\mu\nu}+h_{\mu\nu}) + \frac{1}{2}\left( 4\frac{\partial_{y}a}{a}+d\frac{\partial_{y}b}{b}\right) \partial_{y} h_{\mu\nu} +\frac{1}{2} \frac{\partial_{y}a}{a} \eta_{\mu\nu} \partial_{y} h \nonumber \right. \\ &\left.  +\frac{1}{2} \partial_{y} \partial_{y} h_{\mu\nu}+\frac{1}{2}\frac{1}{b^2}\hat{\Delta}^{(d)} h_{\mu \nu}\right]+\frac{1}{2}\left(\partial_{\lambda} \partial_{\mu} h_{\nu}^{\lambda}+\partial_{\lambda} \partial_{\nu} h^{\lambda}_{\mu}-\square^{(4)} h_{\mu \nu}-\partial_{\nu} \partial_{\mu} h\right)\,, \\
\tilde{R}_{\mu y}=&\frac{1}{2} \partial_{y}\left(\partial_{\lambda} h_{\mu}^{\lambda}-\partial_{\mu} h\right)\,,\\
\tilde{R}_{\mu i}=&\frac{1}{2} \partial_{i}\left(\partial_{\lambda} h_{\mu}^{\lambda}-\partial_{\mu} h\right)\,,\\
\tilde{R}_{yy}=&-\left(4\frac{\partial_{y}\partial_{y}a}{a}+d\frac{\partial_{y}\partial_{y}b}{b}\right)-\frac{1}{2} \partial_{y} \partial_{y} h-\frac{\partial_{y}a}{a} \partial_{y} h\,,\\
\tilde{R}_{yi}=&-\frac{1}{2}\left(\frac{\partial_{y}a}{a}-\frac{\partial_{y}b}{b}\right) \partial_{i} h-\frac{1}{2} \partial_{y} \partial_{i} h\,,\\
\tilde{R}_{ij}=&-b^2\left[\left((d-1)\frac{(\partial_{y}b)^2}{b^2}+\frac{\partial_{y}\partial_{y}b}{b}+4 \frac{\partial_{y}a\partial_{y}b}{ab}\right)+\frac{1}{2} \frac{1}{b^2}\partial_{i} \partial_{j} h+\frac{1}{2} \frac{\partial_{y}b}{b}\delta_{ij} \partial_{y} h\right]\,,
\end{align}
\end{subequations}
where $\square^{(4)}=\eta^{\mu\nu}\partial_\mu\partial_\nu$ and $\hat{\Delta}^{(d)}=\delta^{ij}\partial_i\partial_j$ are the d'Alembert operator in the $\mathcal{M}_4$ and the Laplace operator in the $\mathcal{E}_d$, respectively. The perturbed curvature scalar is
\begin{align}
\tilde{R}=&-\left[4\left(3\frac{(\partial_{y}a)^2}{a^2}+ 2 \frac{\partial_{y}\partial_{y}a}{a}\right)
+d\left((d-1)\frac{(\partial_{y}b)^2}{b^2}+2\frac{\partial_{y}\partial_{y}b}{b}\right)+8d\frac{\partial_{y}a\partial_{y}b}{ab}\right] \nonumber \\ & -\left(5 \frac{\partial_{y}a}{a}+d\frac{\partial_{y}b}{b}\right) \partial_{y} h-\partial_{y} \partial_{y} h- \frac{1}{b^2} \hat{\Delta}^{(d)} h+\frac{1}{a^2}\left(\partial_{\lambda} \partial_{\mu} h^{\lambda \mu}-\square^{(4)} h\right)\,.
\end{align}
The nonvanishing components of the perturbation of the Einstein tensor are
\begin{subequations}
\begin{align}
\tilde{G}_{\mu \nu}=&a^2\left\{\left[3\left(\frac{(\partial_{y}a)^2}{a^2}+\frac{\partial_{y}\partial_{y}a}{a}\right)
+d\left(\frac{(d-1)}{2}\frac{(\partial_{y}b)^2}{b^2}+\frac{\partial_{y}b}{b}\right)
+3d\frac{\partial_{y}a\partial_{y}b}{ab}\right](\eta_{\mu \nu}+h_{\mu \nu}) \nonumber \right.\\ &\left. +\frac{1}{2}\left[\left(4 \frac{\partial_{y}\partial_{y}a}{a}
+d \frac{\partial_{y}b}{b}\right)\partial_{y} h+
\partial_{y} \partial_{y} h+ \frac{1}{b^2} \hat{\Delta}^{(d)} h-\frac{1}{a^2} \left(\partial_{\lambda} \partial_{\mu} h^{\lambda \mu}-\square^{(4)} h\right)\right]\eta_{\mu \nu}\nonumber \right.\\ &\left.
-\frac{1}{2}\left[\left(4 \frac{\partial_{y}a}{a}
+d\frac{\partial_{y}b}{b}\right) \partial_{y} h_{\mu \nu}+\partial_{y} \partial_{y} h_{\mu \nu}
+\frac{1}{b^2} \hat{\Delta}^{(d)} h_{\mu \nu} \nonumber \right.\right.\\ &\left.\left. -\frac{1}{a^2}\left(\partial_{\lambda} \partial_{\mu} h_{\nu}^{\lambda}
+\partial_{\lambda} \partial_{\nu} h^{\lambda}_{\mu}-\square^{(4)} h_{\mu \nu}
-\partial_{\nu} \partial_{\mu} h\right)\right]\right\}\,,\\
\tilde{G}_{\mu y}=&\frac{1}{2} \partial_{y}\left(\partial_{\lambda} h_{\mu}^{\lambda}
-\partial_{\mu} h\right)\,,\\
\tilde{G}_{\mu i}=&\frac{1}{2} \partial_{i}\left(\partial_{\lambda} h_{\mu}^{\lambda}
-\partial_{\mu} h\right)\,,\\
\tilde{G}_{yy}=&6 \frac{(\partial_{y}a)^2}{a^2}+\frac{d}{2}(d-1)\frac{(\partial_{y}b)^2}{b^2}
+4d\frac{\partial_{y}a\partial_{y}b}{ab} \nonumber \\ & +\frac{1}{2}\left(3 \frac{\partial_{y}a}{a}
-d\frac{\partial_{y}b}{b}\right) \partial_{y} h+\frac{1}{2} \frac{1}{b^2} \hat{\Delta}^{(d)}h
-\frac{1}{2} \frac{1}{a^2} \left(\partial_{\lambda} \partial_{\mu} h^{\lambda \mu}-\square^{(4)} h\right)\,,\\
\tilde{G}_{yi}=&-\frac{1}{2}\left(\frac{\partial_{y}a}{a}-\frac{\partial_{y}b}{b}\right) \partial_{i} h
-\frac{1}{2} \partial_{y} \partial_{i} h\,,\\
\tilde{G}_{ij}=& b^2 \left\{\left[2\left(3\frac{(\partial_{y}a)^2}{a^2}
+\frac{\partial_{y}\partial_{y}a}{a}\right)+(d-1)\left(\frac{(d-2)}{2}\frac{(\partial_{y}b)^2}{b^2}
+\frac{\partial_{y}\partial_{y}b}{b}\right)+4(d-1)\frac{\partial_{y}a\partial_{y}b}{ab} \nonumber \right.\right.\\ &\left.\left. +\frac{1}{2}\left(5\frac{\partial_{y}a}{a}+(d-1)\frac{\partial_{y}b}{b}\right)\partial_{y} h
+\frac{1}{2}\partial_{y} \partial_{y} h
-\frac{1}{2}\frac{1}{a^2}\left(\partial_{\lambda} \partial_{\mu} h^{\lambda \mu}-\square^{(4)} h\right)
+\frac{1}{2}\frac{1}{b^2}\hat{\Delta}^{(d)}h\right]\delta_{ij} \nonumber \right.\\ &\left. -\frac{1}{2}\frac{1}{b^2}\partial_i\partial_j h\right\} \,.
\end{align}
\end{subequations}

Taking into account the condition~\eqref{TTcondition}, we list all components of the curvature scalar, the Ricci tensor, and the Einstein tensor only in the linear order perturbation in the following. All components of the perturbed Ricci tensor are simplified as
\begin{align}\label{TTRMN}
\delta R_{\mu\nu}=&a^2\left\{-\frac{1}{2}\left[\left( 4\frac{\partial_{y}a}{a}+d\frac{\partial_{y}b}{b}\right) \partial_{y} h_{\mu\nu} + \partial_{y} \partial_{y} h_{\mu\nu} +\frac{1}{a^2}\square^{(4)}h_{\mu\nu}+\frac{1}{b^2}\hat{\Delta}^{(d)} h_{\mu \nu} \right]\nonumber \right. \\ &\left. -\left(3\frac{(\partial_{y}a)^2}{a^2}
+\frac{(\partial_{y}a)^2}{a}+d\frac{\partial_{y}a\partial_{y}b}{ab}\right)h_{\mu\nu}\right\}\,, \\
\delta R_{\mu y}=&0\,,\quad \delta R_{\mu i}=0\,,\quad \delta R_{yy}=0\,,\quad \delta R_{yi}=0\,, \quad \delta R_{ij}=0\,, \nonumber
\end{align}
and the perturbed curvature scalar vanishes
\begin{align}\label{TTR}
  \delta R=&0\,.
\end{align}
All components of the perturbed Einstein tensor are given by
\begin{align}
  \delta G_{\mu \nu} =&a^2\left\{-\frac{1}{2}\left[\left(4 \frac{\partial_{y}a}{a}+d\frac{\partial_{y}b}{b}\right) \partial_{y} h_{\mu \nu}+\partial_{y} \partial_{y} h_{\mu \nu}+\frac{1}{a^2}\square^{(4)} h_{\mu \nu}+\frac{1}{b^2} \hat{\Delta}^{(d)} h_{\mu \nu} \right] \nonumber \right.\\ &\left. +\left[3\left(\frac{(\partial_{y}a)^2}{a^2}+\frac{(\partial_{y}a)^2}{a}\right)
  +d\left(\frac{(d-1)}{2}\frac{(\partial_{y}b)^2}{b^2}+\frac{\partial_{y}\partial_{y}b}{b}\right)
  +3d\frac{\partial_{y}a\partial_{y}b}{ab}\right]h_{\mu \nu}\right\} \,, \label{deltaGmunu} \\
  \delta G_{\mu y} =&0\,, \quad \delta G_{\mu i} =0\,, \quad \delta G_{y y} =0 \,, \quad \delta G_{y i} =0 \,, \quad \delta G_{ij} =0 \,. \nonumber
\end{align}
It is worth noting that the $\delta$ used here merely stands for linear order quantities under perturbations.

\section{Two Linearly Independent Solutions of the Schr\"{o}dinger-like Equation}
\label{appendix3}

We present some details about two linearly independent (nontrivial) solutions for the Schr\"{o}dinger-like equation~\eqref{eq:Schrodinger} or~\eqref{eq:ssform} with zero eigenvalue, namely
\begin{equation}\label{eq:zeroenergy}
 \mathcal{Q}\,\mathcal{Q}^{\dagger}\psi(z)=0\,.
\end{equation}
Using the following equation
\begin{align}\label{eq:Qdagger}
  (-\partial_z+\Omega)\psi(z)=0\,,
\end{align}
we obtain a solution as
\begin{align}\label{solution_1}
  \psi_1(z)=C_1\left[a^3(z)b^d(z)f_{R}(z)\right]^{1/2}\,,
\end{align}
where $C_1$ is an integration constant. Moreover, for the following equation
\begin{align}\label{eq:Q}
  (\partial_z+\Omega)\check{\psi}(z)=0\,,
\end{align}
the corresponding solution is given by
\begin{align}\label{solutionprime}
  \psi_c(z)=C^{\prime} \left[a^3(z)b^d(z)f_{R}(z)\right]^{-1/2}\,.
\end{align}
where $C^{\prime}$ is an integration constant. If we require Eq.~\eqref{eq:zeroenergy} always holds, we will arrive at
\begin{align}
  (-\partial_z+\Omega)\psi(z)=\psi_c(z)\,.
\end{align}
Therefore, another solution for Eq.~\eqref{eq:zeroenergy} is
\begin{align}
  \psi_2(z)=C_2\left[a^3(z)b^d(z)f_{R}(z)\right]^{1/2}\int
  \frac{1}{a^3(z)b^d(z)f_{R}(z)}\mathrm{d}z\,,
\end{align}
where $C_2$ is an integration constant. The general (nontrivial) solution for Eq.~\eqref{eq:zeroenergy} is given by
\begin{align}
  \psi(z)=\left[a^3(z)b^d(z)f_{R}(z)\right]^{1/2}\left[C_1+ C_2 \int
  \frac{1}{a^3(z)b^d(z)f_{R}(z)}\mathrm{d}z\right]\,.
\end{align}


\begin{thebibliography}{}



\bibitem{Sotiriou:2010sf}
 T. P. Sotiriou and V. Faraoni,
    \emph{$f(R)$ theories of gravity},
    \href{https://dx.doi.org/10.1103/RevModPhys.82.451}{Rev. Mod. Phys. \textbf{82} (2010) 451},
    \href{https://arxiv.org/abs/0805.1726}{arXiv:0805.1726 [gr-qc]}.

\bibitem{De Felice:2010dt}
 A. De Felice and S. Tsujikawa,
    \emph{$f(R)$ theories},
    \href{https://dx.doi.org/10.12942/lrr-2010-3}{Living Rev. Rel. \textbf{13} (2010)},
    \href{https://arxiv.org/abs/1002.4928}{arXiv:1002.4928 [gr-qc]}.

\bibitem{Randall:1999rsa}
 L. Randall and R. Sundrum,
    \emph{An alternative to compactification},
    \href{https://dx.doi.org/10.1103/PhysRevLett.83.4690}{Phys. Rev. Lett. \textbf{83} (1999) 4690},
    \href{https://arxiv.org/abs/hep-th/9906064}{arXiv:hep-th/9906064}.

\bibitem{da Silva:2011sd}
 J. M. Hoff da Silva and M. Dias,
    \emph{Five-dimensional $f(R)$ braneworld models},
    \href{https://dx.doi.org/10.1103/PhysRevD.84.066011}{Phys. Rev. \textbf{D 84} (2011) 066011},
    \href{https://arxiv.org/abs/1107.2017}{arXiv:1107.2017 [hep-th]}.

\bibitem{Parry:2005ppd}
 M. Parry, S. Pichler, and D. Deeg,
    \emph{Higher-derivative gravity in brane world models},
    \href{https://dx.doi.org/10.1088/1475-7516/2005/04/014}{JCAP \textbf{04} (2005) 014},
    \href{https://arxiv.org/abs/hep-ph/0502048}{arXiv:hep-ph/0502048}.

\bibitem{Balcerzak:2008bd}
 A. Balcerzak and M. P. Dabrowski,
    \emph{Generalized Israel junction conditions for a fourth-order brane world},
    \href{https://dx.doi.org/10.1103/PhysRevD.77.023524}{Phys. Rev. \textbf{D 77} (2008) 023524},
    \href{https://arxiv.org/abs/0710.3670}{arXiv:0710.3670 [hep-th]}.

\bibitem{Deruelle:2007dss}
 N. Deruelle, M. Sasaki, and Y. Sendouda,
    \emph{Junction conditions in $f(R)$ theories of gravity},
    \href{https://dx.doi.org/10.1143/PTP.119.237}{Prog. Theor. Phys. \textbf{119} (2008) 237},
    \href{https://arxiv.org/abs/0711.1150}{arXiv:0711.1150 [gr-qc]}.

\bibitem{Balcerzak:2009bd}
 A. Balcerzak and M. Dabrowski,
    \emph{Gibbons-Hawking boundary terms and junction conditions for higher-order brane gravity models},
    \href{https://dx.doi.org/10.1088/1475-7516/2009/01/018}{JCAP \textbf{01} (2009) 018},
    \href{https://arxiv.org/abs/0804.0855}{arXiv:0804.0855 [hep-th]}.

\bibitem{Borzou:2009bssy}
 A. Borzou, H. R. Sepangi, S. Shahidi, and R. Yousefi,
    \emph{Brane $f(R)$ gravity},
    \href{https://dx.doi.org/10.1209/0295-5075/88/29001}{Europhys. Lett. \textbf{88} (2009) 29001},
    \href{https://arxiv.org/abs/0910.1933}{arXiv:0910.1933 [gr-qc]}.

\bibitem{Balcerzak:2011bd}
 A. Balcerzak and M. P. Dabrowski,
    \emph{Randall-Sundrum limit of $f(R)$ brane-world models},
    \href{https://dx.doi.org/10.1103/PhysRevD.84.063529}{Phys. Rev. \textbf{D 84} (2011) 063529},
    \href{https://arxiv.org/abs/1107.3048}{arXiv:1107.3048 [hep-th]}.

\bibitem{Carames:2013cgs}
 T. R. P. Caram\^{e}s, M. E. X. Guimar\~{a}es, and J. M. Hoff da Silva,
    \emph{Effective gravitational equations for $f(R)$ braneworld models},
    \href{https://dx.doi.org/10.1103/PhysRevD.87.106011}{Phys. Rev. \textbf{D 87} (2013) 106011},
    \href{https://arxiv.org/abs/1205.4980}{arXiv:1205.4980 [gr-qc]}.

\bibitem{Rubakov:1983rsd}
 V. A. Rubakov and M. E. Shaposhnikov,
    \emph{Do we live inside a domain wall?},
    \href{https://dx.doi.org/10.1016/0370-2693(83)91253-4}{Phys. Lett. \textbf{B 125} (1983) 136}.

\bibitem{Dzhunushaliev:2010dfm}
 V. Dzhunushaliev, V. Folomeev, and M. Minamitsuji,
    \emph{Thick brane solutions},
    \href{https://dx.doi.org/10.1088/0034-4885/73/6/066901}{Rept. Prog. Phys. \textbf{73} (2010) 066901},
    \href{https://arxiv.org/abs/0904.1775}{arXiv:0904.1775 [gr-qc]}.

\bibitem{Maartens:2010gt}
 R. Maartens and K. Koyama,
    \emph{Brane-world gravity},
    \href{https://dx.doi.org/10.12942/lrr-2010-5}{Living Rev. Rel. \textbf{13} (2010) 5},
    \href{https://arxiv.org/abs/1004.3962}{arXiv:1004.3962 [hep-th]}.

\bibitem{Liu:2017l}
 Y.-X. Liu,
    \emph{Introduction to extra dimensions and thick braneworlds},
    \href{https://dx.doi.org/10.1142/9789813237278_0008}{Memorial Volume for Yi-Shi Duan (2018) 211},
    \href{https://arxiv.org/abs/1707.08541}{arXiv:1707.08541 [hep-th]}.

\bibitem{Zhong:2016zl}
 Y. Zhong and Y.-X. Liu,
    \emph{Pure geometric thick $f(R)$-branes: stability and localization of gravity},
    \href{https://dx.doi.org/10.1140/epjc/s10052-016-4163-0}{Eur. Phys. J. \textbf{C 76} (2016) 321},
    \href{https://arxiv.org/abs/1507.00630}{arXiv:1507.00630 [hep-th]}.

\bibitem{Dzhunushaliev:2010dfkk}
 V. Dzhunushaliev, V. Folomeev, B. Kleihaus, and J. Kunz,
    \emph{Some thick brane solutions in $f(R)$-gravity},
    \href{https://dx.doi.org/10.1007/JHEP04(2010)130}{JHEP \textbf{04} (2010) 130},
    \href{https://arxiv.org/abs/0912.2812}{arXiv:0912.2812 [gr-qc]}.

\bibitem{Dzhunushaliev:2019dfno}
 V. Dzhunushaliev, V. Folomeev, G. Nurtaeva, and S. D. Odintsov,
    \emph{Thick branes in higher-dimensional $f(R)$ gravity},
    \href{https://dx.doi.org/10.1142/S021988782050036X}{Int. J. Geom. Meth. Mod. Phys. \textbf{17} (2020) 2050036},
    \href{https://arxiv.org/abs/1908.01312}{arXiv:1908.01312 [gr-qc]}.

\bibitem{Dzhunushaliev:2019dfs}
 V. Dzhunushaliev, V. Folomeev, and A. Serikbolova,
    \emph{Codimension-$1$ thick brane solutions in higher-dimensional $R^n$ gravity},
    \href{https://arxiv.org/abs/1912.13395}{arXiv:1912.13395 [gr-qc]}.

\bibitem{Afonso:2007abmp}
 V. I. Afonso, D. Bazeia, R. Menezes, and A. Yu. Petrov,
    \emph{$f(R)$-brane},
    \href{https://dx.doi.org/10.1016/j.physletb.2007.10.038}{Phys. Lett. \textbf{B 658} (2007) 71},
    \href{https://arxiv.org/abs/0710.3790}{arXiv:0710.3790 [hep-th]}.

\bibitem{Liu:2011lzzl}
 Y.-X. Liu, Y. Zhong, Z.-H. Zhao, and H.-T. Li,
    \emph{Domain wall brane in squared curvature gravity},
    \href{https://dx.doi.org/10.1007/JHEP06(2011)135}{JHEP \textbf{06} (2011) 135},
    \href{https://arxiv.org/abs/1104.3188}{arXiv:1104.3188 [hep-th]}.

\bibitem{Liu:2012llw}
 H. Liu, H. L\"{u}, and Z.-L. Wang,
    \emph{$f(R)$ gravities, Killing spinor equations, ``BPS'' domain walls and cosmology},
    \href{https://dx.doi.org/10.1007/jhep02(2012)083}{JHEP \textbf{02} (2012) 83},
    \href{https://arxiv.org/abs/1111.6602}{arXiv:1111.6602 [hep-th]}.

\bibitem{Bazeia:2013bmps}
 D. Bazeia, R. Menezes, A. Yu. Petrov, and  A. J. da Silva,
    \emph{On the many-field $f(R)$ brane},
    \href{https://dx.doi.org/10.1016/j.physletb.2013.08.068}{Phys. Lett. \textbf{B 726} (2013) 523},
    \href{https://arxiv.org/abs/1306.1847}{arXiv:1306.1847 [hep-th]}.

\bibitem{Bazeia:2014blmps}
 D. Bazeia, A. S. Lob\~ao Jr., R. Menezes, A. Yu. Petrov, and  A. J. da Silva,
    \emph{Braneworld solutions for $F(R)$ models with non-constant curvature},
    \href{https://dx.doi.org/10.1016/j.physletb.2014.01.011}{Phys. Lett. \textbf{B 729} (2014) 127},
    \href{https://arxiv.org/abs/1311.6294}{arXiv:1311.6294 [hep-th]}.

\bibitem{Xu:2015xzyl}
 Z.-G. Xu, Y. Zhong, H. Yu, and Y.-X. Liu,
    \emph{The structure of $f(R)$-brane model},
    \href{https://dx.doi.org/10.1140/epjc/s10052-015-3597-0}{Eur. Phys. J. \textbf{C 75} (2015) 368},
    \href{https://arxiv.org/abs/1405.6277}{arXiv:1405.6277 [hep-th]}.

\bibitem{Bazeia:2014blmor}
 D. Bazeia, L. Losano, R. Menezes, G. J. Olmo, and D. Rubiera-Garcia,
    \emph{Thick brane in $f(R)$ gravity with Palatini dynamics},
    \href{https://dx.doi.org/10.1140/epjc/s10052-015-3803-0}{Eur. Phys. J. \textbf{C 75} (2015) 569},
    \href{https://arxiv.org/abs/1411.0897}{arXiv:1411.0897 [hep-th]}.

\bibitem{Gu:2015ggyl}
 B.-M. Gu, B. Guo, H. Yu, and Y.-X. Liu,
    \emph{Tensor perturbations of Palatini $f(R)$-branes},
    \href{https://dx.doi.org/10.1103/PhysRevD.92.024011}{Phys. Rev. \textbf{D 92} (2015) 024011},
    \href{https://arxiv.org/abs/1411.3241}{arXiv:1411.3241 [hep-th]}.

\bibitem{Bazeia:2015blm}
 D. Bazeia, A. S. Lob\~ao Jr., and R. Menezes,
    \emph{Thick brane models in generalized theories of gravity},
    \href{https://dx.doi.org/10.1016/j.physletb.2015.02.037}{Phys. Lett. \textbf{B 743} (2015) 98},
    \href{https://arxiv.org/abs/1502.04757}{arXiv:1502.04757 [hep-th]}.

\bibitem{Bazeia:2015bllmo}
 D. Bazeia, A. S. Lob\~ao Jr., L. Losano, R. Menezes, and G. J. Olmo,
    \emph{Braneworld solutions for modified theories of gravity with nonconstant curvature},
    \href{https://dx.doi.org/10.1103/PhysRevD.91.124006}{Phys. Rev. \textbf{D 91} (2015) 124006},
    \href{https://arxiv.org/abs/1505.06315}{arXiv:1505.06315 [hep-th]}.

\bibitem{Gu:2017gzyl}
 B.-M. Gu, Y.-P. Zhang, H. Yu, and Y.-X. Liu,
    \emph{Full linear perturbations and localization of gravity on $f(R,T)$ brane},
    \href{https://dx.doi.org/10.1140/epjc/s10052-017-4666-3}{Eur. Phys. J. \textbf{C 77} (2017) 115},
    \href{https://arxiv.org/abs/1606.07169}{arXiv:1606.07169 [hep-th]}.

\bibitem{Cui:2018clgz}
 Z.-Q. Cui, Y.-X. Liu, B.-M. Gu, and L. Zhao,
    \emph{Linear stability of $f(R,\phi,X)$ thick branes: tensor perturbations},
    \href{https://dx.doi.org/10.1007/JHEP11(2018)083}{JHEP \textbf{11} (2018) 083},
    \href{https://arxiv.org/abs/1802.01454}{arXiv:1802.01454 [hep-th]}.

\bibitem{Gu:2018glz}
 B.-M. Gu, Y.-X. Liu, and Y. Zhong,
    \emph{Stable Palatini $f(R)$ braneworld},
    \href{https://dx.doi.org/10.1103/PhysRevD.98.024027}{Phys. Rev. \textbf{D 98} (2018) 024027},
    \href{https://arxiv.org/abs/1804.00271}{arXiv:1804.00271 [hep-th]}.


\bibitem{Wang:2019wgfx}
 L.-L. Wang, H. Guo, C.-E. Fu, and Q.-Y. Xie,
    \emph{Gravity and matters on a pure geometric thick polynomial $f(R)$ brane},
    \href{https://arxiv.org/abs/1912.01396}{arXiv:1912.01396 [hep-th]}.

\bibitem{Yu:2016yzgl}
 H. Yu, Y. Zhong, B.-M. Gu, and Y.-X. Liu,
    \emph{Gravitational resonances on $f(R)$-brane},
    \href{https://dx.doi.org/10.1140/epjc/s10052-016-4039-3}{Eur. Phys. J. \textbf{C 76} (2016) 195},
    \href{https://arxiv.org/abs/1506.06458}{arXiv:1506.06458 [gr-qc]}.

\bibitem{Giovannini:1997g}
 M. Giovannini,
    \emph{Scalar and tensor inhomogeneities from dimensional decoupling},
    \href{https://dx.doi.org/10.1103/PhysRevD.55.595}{Phys. Rev. \textbf{D 55} (1997) 595},
    \href{https://arxiv.org/abs/hep-th/9610179}{arXiv:hep-th/9610179}.

\bibitem{Csaki:2000cehs}
 C. Cs\'{a}ki, J. Erlich, T. J. Hollowood, and Y. Shirman,
    \emph{Universal aspects of gravity localized on thick branes},
    \href{https://dx.doi.org/10.1016/S0550-3213(00)00271-6}{Nucl. Phys. \textbf{B 581} (2000) 309},
    \href{https://arxiv.org/abs/hep-th/0001033}{arXiv:hep-th/0001033}.

\bibitem{Giovannini:2001gg}
 M. Giovannini,
    \emph{Gauge-invariant fluctuations of scalar branes},
    \href{https://dx.doi.org/10.1103/PhysRevD.64.064023}{Phys. Rev. \textbf{D 64} (2001) 064023},
    \href{https://arxiv.org/abs/hep-th/0106041}{arXiv:hep-th/0106041}.

\bibitem{Giovannini:2002ga}
 M. Giovannini,
    \emph{Localization of metric fluctuations on scalar branes},
    \href{https://dx.doi.org/10.1103/PhysRevD.65.064008}{Phys. Rev. \textbf{D 65} (2002) 064008},
    \href{https://arxiv.org/abs/hep-th/0106131}{arXiv:hep-th/0106131}.

\bibitem{Kobayashi:2002kks}
 S. Kobayashi, K. Koyama, and J. Soda,
    \emph{Thick brane worlds and their stability},
    \href{https://dx.doi.org/10.1103/PhysRevD.65.064014}{Phys. Rev. \textbf{D 65} (2002) 064014},
    \href{https://arxiv.org/abs/hep-th/0107025}{arXiv:hep-th/0107025}.

\bibitem{Giovannini:2002gb}
 M. Giovannini,
    \emph{Theory of gravitational fluctuations in brane world models},
    \href{https://dx.doi.org/10.1142/s0218271802002177}{Int. J. Mod. Phys. \textbf{D 11} (2002) 1209}.

\bibitem{Giovannini:2003g}
 M. Giovannini,
    \emph{Scalar normal modes of higher-dimensional gravitating kinks},
    \href{https://dx.doi.org/10.1088/0264-9381/20/6/303}{Class. Quant. Grav. \textbf{20} (2003) 1063},
    \href{https://arxiv.org/abs/gr-qc/0207116}{arXiv:gr-qc/0207116}.

\bibitem{Zhong:2011zly}
 Y. Zhong, Y.-X. Liu, and K. Yang,
    \emph{Tensor perturbations of $f(R)$-branes},
    \href{https://dx.doi.org/10.1016/j.physletb.2011.04.037}{Phys. Lett. \textbf{B 699} (2011) 398},
    \href{https://arxiv.org/abs/1010.3478}{arXiv:1010.3478 [hep-th]}.

\bibitem{Zhong:2017zl}
 Y. Zhong and Y.-X. Liu,
    \emph{Linearization of a warped $f(R)$ theory in the higher-order frame},
    \href{https://dx.doi.org/10.1103/PhysRevD.95.104060}{Phys. Rev. \textbf{D 95} (2017) 104060},
    \href{https://arxiv.org/abs/1611.08237}{arXiv:1611.08237 [gr-qc]}.

\bibitem{Chen:2017cgl}
 F.-W. Chen, B.-M. Gu, and Y.-X. Liu,
    \emph{Stability of braneworlds with non-minimally coupled multi-scalar fields},
    \href{https://dx.doi.org/10.1140/epjc/s10052-018-5613-7}{Eur. Phys. J. \textbf{C 78} (2018) 131},
    \href{https://arxiv.org/abs/1702.03497}{arXiv:1702.03497 [hep-th]}.

\bibitem{Gherghetta:2010}
 T. Gherghetta,
    \emph{A holographic view of beyond the standard model physics}, in
    \href{https://dx.doi.org/10.1142/9789814327183_0004}{\emph{Theoretical Advanced Study Institute in Elementary Particle Physics: Physics of the Large and the Small} (2011) 165},
    \href{https://arxiv.org/abs/1008.2570}{arXiv:1008.2570 [hep-ph]}.


\bibitem{Bronnikov:2017bpr}
 K. A. Bronnikov, R. I. Budaev, A. V. Grobov, A. E. Dmitriev and S. G. Rubin,
    \emph{Inhomogeneous compact extra dimensions},
    \href{https://dx.doi.org/10.1088/1475-7516/2017/10/001}{JCAP \textbf{10} (2017) 001},
    \href{https://arxiv.org/abs/1707.00302}{arXiv:1707.00302 [gr-qc]}.

\bibitem{Bronnikov:2020bpr}
 K. A. Bronnikov, A. A. Popov, and S. G. Rubin,
    \emph{Inhomogeneous compact extra dimensions and de Sitter cosmology},
    \href{https://dx.doi.org/10.1140/epjc/s10052-020-08547-x}{Eur. Phys. J. \textbf{C 80} (2020) 970},
    \href{https://arxiv.org/abs/2004.03277}{arXiv:2004.03277 [gr-qc]}.

\bibitem{Starobinsky:1980s}
 A. Starobinsky,
    \emph{A new type of isotropic cosmological models without singularity},
    \href{https://dx.doi.org/10.1016/0370-2693(80)90670-X}{Phys. Lett. \textbf{B 91} (1980) 99}.

\bibitem{Higuchi:2014hn}
 M. Higuchi and S. Nojiri,
    \emph{Reconstruction of domain wall universe and localization of gravity},
    \href{https://dx.doi.org/10.1007/s10714-014-1822-z}{Gen. Rel. Grav. \textbf{46} (2014) 1822},
    \href{https://arxiv.org/abs/1402.1346}{arXiv:1402.1346 [hep-th]}.

\bibitem{Chakraborty:2016cs}
 S. Chakraborty and S. SenGupta,
    \emph{Solving higher curvature gravity theories},
    \href{https://dx.doi.org/10.1140/epjc/s10052-016-4394-0}{Eur. Phys. J. \textbf{C 76} (2016) 552},
    \href{https://arxiv.org/abs/1604.05301}{arXiv:1604.05301 [gr-qc]}.

\bibitem{Chakraborty:2015csfr}
 S. Chakraborty and S. SenGupta,
    \emph{Spherically symmetric brane spacetime with bulk $f(R)$ gravity},
    \href{https://dx.doi.org/10.1140/epjc/s10052-014-3234-3}{Eur. Phys. J. \textbf{C 75} (2015) 11},
    \href{https://arxiv.org/abs/1409.4115}{arXiv:1409.4115 [gr-qc]}.

\bibitem{Chakraborty:2016csfrgb}
 S. Chakraborty and S. SenGupta,
    \emph{Spherically symmetric brane in a bulk of $f(R)$ and Gauss–Bonnet gravity},
    \href{https://dx.doi.org/10.1088/0264-9381/33/22/225001}{Class. Quant. Grav. \textbf{33} (2016) 225001},
    \href{https://arxiv.org/abs/1510.01953}{arXiv:1510.01953 [gr-qc]}.

\bibitem{Zhou:2018zdyl}
 X.-N. Zhou, Y.-Z. Du, H. Yu, and Y.-X. Liu,
    \emph{Localization of gravitino field on $f(R)$ thick branes},
    \href{https://dx.doi.org/10.1007/s11433-018-9246-2}{Sci. China Phys. Mech. Astron. \textbf{61} (2018) 110411},
    \href{https://arxiv.org/abs/1703.10805}{arXiv:1703.10805 [hep-th]}.

\bibitem{Mitra:2017mps}
 J. Mitra, T. Paul, and S. SenGupta,
    \emph{Fermion localization in higher curvature and scalar-tensor theories of gravity},
    \href{https://dx.doi.org/10.1140/epjc/s10052-017-5420-6}{Eur. Phys. J. \textbf{C 77} (2017) 833},
    \href{https://arxiv.org/abs/1707.06532}{arXiv:1707.06532 [hep-th]}.






	




\end{thebibliography}

\end{document}